\documentclass[sigconf]{acmart}

\AtBeginDocument{%
  \providecommand\BibTeX{{%
    \normalfont B\kern-0.5em{\scshape i\kern-0.25em b}\kern-0.8em\TeX}}}

\setcopyright{rightsretained}

\acmConference[MICRO'19]{The 52nd Annual IEEE/ACM International Symposium on Microarchitecture}{October 12-16, 2019}{Columbus, OH, USA}
\acmConference[]{}{}{}
\acmDOI{10.1145/XXXXX.XXXXXX}
\acmISBN{}

\usepackage{comment}
\usepackage{mathtools}
\usepackage{setspace}
\usepackage{algorithm}
\usepackage[noend]{algorithmic}
\usepackage{graphics}
\usepackage{fancyhdr}
\usepackage{subcaption}
\usepackage{booktabs}
\usepackage{array}
\usepackage[normalem]{ulem}
\makeatletter
\def\mdseries@tt{m}
\makeatother
\usepackage[newfloat,frozencache]{minted}

\usepackage{hhline}
\usepackage{caption}
\usepackage{tikz}
\usepackage{xcolor}
\usepackage{mdframed}
\usepackage{hyperref}
\usepackage{breakurl}
\usepackage{makecell}
\usepackage{textcomp}
\usepackage{dblfloatfix}
\usepackage{hhline}
\usepackage{float}
\usepackage[inline]{enumitem}
\usepackage{afterpage}
\usepackage{graphicx}
\usepackage{listings}
\definecolor{dkgreen}{rgb}{0,0.6,0}
\definecolor{gray}{rgb}{0.5,0.5,0.5}
\definecolor{mauve}{rgb}{0.58,0,0.82}
\definecolor{chocolate}{rgb}{0.48, 0.25, 0.0}

\newcommand*\circled[1]{\tikz[baseline=(char.base)]{
            \node[shape=circle,fill,inner sep=0.5pt] (char) {\textcolor{white}{#1}};}}

\newcommand{\myname}{SMASH}
\newcommand\kon[1]{\noindent{\color{black}#1}} 

\newenvironment{code}{\captionsetup{type=listing}}{}
\SetupFloatingEnvironment{listing}{name=Code Listing}
\begin{document}

\acmConference[MICRO-52]{The 52nd Annual IEEE/ACM International Symposium on Microarchitecture}{October 12--16, 2019}{Columbus, OH, USA}
\acmDOI{10.1145/3352460.3358286}
\title{ SMASH: Co-designing Software Compression \\ and Hardware-Accelerated  Indexing  \\ for Efficient Sparse Matrix Operations}

\renewcommand{\shorttitle}{SMASH: Co-designing Software Compression and Hardware-Accelerated  Indexing \\ for Efficient Sparse Matrix Operations}




\author{Konstantinos Kanellopoulos$^1$ \space  Nandita Vijaykumar$^{2,1}$ \space   Christina Giannoula$^{1,3}$  \space   Roknoddin Azizi$^1$}
\author{Skanda Koppula$^1$ \space Nika Mansouri Ghiasi$^1$   \space Taha Shahroodi$^1$ \space Juan Gomez Luna$^1$ \space  Onur Mutlu$^{1,2}$}

\affiliation{  \vspace{0.15cm} $^1$ETH Z{\"u}rich  \space  \space  \space  \space  \space  \space     $^2$Carnegie Mellon University  \space  \space  \space  \space  \space  \space     $^3$National Technical University of Athens}

\renewcommand{\shortauthors}{Kanellopoulos et al.}

\begin{abstract}

Important workloads, such as machine learning and graph analytics applications, heavily involve sparse linear algebra operations.
These operations use sparse matrix compression as an effective means to avoid storing zeros and performing unnecessary computation on zero elements. However, compression techniques like Compressed Sparse Row (CSR) that are widely used today introduce significant instruction overhead and expensive pointer-chasing operations to discover the positions of the non-zero elements. In this paper, we identify the discovery of the positions (i.e., indexing) of non-zero elements as a key bottleneck in sparse matrix-based workloads, which greatly reduces the benefits of compression.


We propose SMASH, a hardware-software cooperative mechanism that enables highly-efficient indexing and storage of sparse matrices. The key idea of SMASH is to explicitly enable the hardware to recognize and exploit sparsity in data. To this end, we devise a novel software encoding based on a \emph{hierarchy of bitmaps}. This encoding can be used to efficiently compress any sparse matrix, regardless of the extent and structure of sparsity. At the same time, the bitmap encoding can be directly interpreted by the hardware. We design a lightweight hardware unit, the Bitmap Management Unit (BMU), that buffers and scans the bitmap hierarchy to perform highly-efficient indexing of sparse matrices. SMASH exposes an expressive and rich ISA to communicate with the BMU, which enables its use in accelerating any sparse matrix computation.

We demonstrate the benefits of SMASH on four use cases that include sparse matrix kernels and graph analytics applications. Our evaluations show that SMASH provides average performance improvements of $38\%$ for Sparse Matrix Vector Multiplication and $44\%$ for Sparse Matrix Matrix Multiplication, over a state-of-\kon{the}-art CSR implementation, on a wide variety of matrices with different characteristics. SMASH incurs a very modest hardware area overhead of up to $0.076\%$ of an out-of-order CPU core.



\end{abstract}
\maketitle

\fancyhead{}

\newif\ifcameraready
\camerareadytrue

\newif\ifwebversion
\webversiontrue

\fancyhead{}
\ifcameraready
  \ifwebversion
    \fancypagestyle{firststyle}
    {
        \setlength{\footskip}{40pt}
        \fancyfoot[C]{\thepage}
    }
    \thispagestyle{firststyle}
    \pagestyle{firststyle}
  \fi
\else
  \fancyhead[C]{\textcolor{teal}{\emph{Version \versionnum~---~\today, \ampmtime}}}
  \fancypagestyle{firststyle}
  {
    \fancyhead[C]{\textcolor{teal}{\emph{Version \versionnum~---~\today, \ampmtime}}}
    \fancyfoot[C]{\thepage}
  }
  \thispagestyle{firststyle}
  \pagestyle{firststyle}
\fi

\setstretch{0.98}

\vspace{-2mm}
\section{Introduction}
\label{Introductionlbl}

Sparse linear algebra operations are widely used in modern applications like recommender systems~\cite{linden2003amazon,recommenderFB2019,recommendfb2}, neural networks~\cite{Benjamin2014spatially, liu2015sparse}, graph analytics~\cite{Brin1998the,besta2017slimsell}, and high-performance computing~\cite{solversGPU, Falgout2006an,dongarra1996sparse, falgout2002hypre, henon2002pastix}. The matrices involved in these operations are very large in size and highly sparse, i.e., the vast majority of the elements are zeros. For example, the matrices that represent Facebook's and YouTube's social network connectivity contain 0.0003\%~\cite{SmithFacebook} and 2.31\%~\cite{leskovec2016snap} non-zero elements, respectively. These highly sparse matrices lead to significant inefficiencies in both storage and computation. First, they require an unnecessarily large amount of storage space, which is largely occupied by zeros. Second, computation on highly sparse matrices involves a large number of unnecessary operations (such as additions and multiplications) on zero elements. The traditional solution to these inefficiencies is to compress the matrix and store only the non-zero elements, and then operate only on the non-zero values.

Prior works take two major approaches to designing such compression schemes. The first approach is to devise general compression \emph{formats} or encodings ~\cite{datacompression2006,im1999optimizing,formatSurvey2016,csr52015,variableblock2005,blockoptim1999,Sengupta2007scan,bccoo2014}, such as CSR~\cite{csr52015}, COO~\cite{Sengupta2007scan}, BCSR~\cite{im1999optimizing}, and CSR5~\cite{csr52015}. Such formats essentially store the non-zero elements and their positions within the matrix using additional data structures and different encodings. Such encodings are general in applicability and are highly-efficient in storage, with high compression ratios. However, this approach involves repositioning and packing the non-zero values in the matrix\kon{,} which leads to significant computation overheads that diminish the overall benefit. Determining the positions of the non-zero elements in the compressed encoding (i.e., \textbf{indexing}) requires a series of pointer-chasing operations in memory that, as we demonstrate, are highly inefficient in modern processors and memory hierarchies. 

The second approach taken by prior work is to leverage a certain known structure in a given type of sparse matrix to avoid the cost of discovering the non-zero regions of the sparse matrix ~\cite{uniquekourtis2008,csxkourtis2011,smatautotune2013,csb2009,pattern2009,cocktail2012,csradaptgpus2014}. For example, the DIA format \cite{pattern2009} is highly effective in matrices where the non-zero values are concentrated along the diagonals of the matrix. Specializing the compression scheme to patterns in the sparsity can be efficient in both computation and storage but it is highly specific to certain types of matrices and inapplicable when the structure and extent of sparsity are not known a priori.

Our \textbf{goal} in this work is to enable efficient and general sparse matrix computation with a technique that satisfies \textbf{three major requirements}: 1) high computation and storage efficiency by storing and operating on only non-zero elements; 2) minimal overheads from the compression scheme (e.g., efficient discovery of non-zero elements) and 3) generality and applicability to any sparse matrix, regardless of its structure or the extent of its sparsity.


Our \textbf{key idea} is a new hardware-software co-design where we enable the hardware to \emph{recognize} and \emph{exploit} the compression encoding used in software for any sparse matrix. This allows us to add hardware support for highly-efficient storage and retrieval of non-zero values in sparse matrices, avoiding the overheads of software indexing (requirement 2). 
Our software encoding is designed to maintain low storage requirements (requirement 1) and be generally applicable to any sparse matrix without any assumption of structure or extent of sparsity (requirement 3). 

We propose SMASH (\textbf{S}parse \textbf{MA}trix \textbf{S}oftware/\textbf{H}ardware), a general hardware-software cooperative mechanism that efficiently compresses and operates on sparse matrices. The key construct behind SMASH is the use of efficiently encoded \emph{hierarchical bitmaps} to express sparsity, where each bit represents a region of non-zero values. These bitmaps are recognized by both hardware and software. 
On the software side, sparse matrices of any form are flexibly encoded using our hierarchical bitmap representation. SMASH adapts to each matrix's sparsity characteristics by supporting multiple compression granularities throughout the bitmap hierarchy. On the hardware side,  the bitmap representation allows us to use a lightweight hardware unit, the Bitmap Management Unit (BMU), to perform highly-efficient scans of the bitmap hierarchy. The BMU hardware enables low-cost indexing (that avoids expensive pointer-chasing lookups) and efficient sparse matrix computation. 

To enable wide applicability, \myname{} exposes five new instructions that enable the software to communicate with the Bitmap Management Unit. The new instructions enable efficient lookup of non-zero matrix regions, and are sufficiently rich to express a wide variety of operations on any type of (sparse) matrix. 

We evaluate \myname{} on four use cases: two sparse matrix kernels, Sparse Matrix Vector Multiplication (SpMV) and  Sparse Matrix-Matrix Multiplication (SpMM), as well as two graph analytics applications, PageRank and Betweenness Centrality (BC). For our experiments, we use a collection of sparse matrices with varying structure and sparsity characteristics ~\cite{florida}.  We compare SMASH to two state-of-the-art compression formats, CSR \cite{csr52015} and BCSR \cite{im1999optimizing}.  We find that SMASH improves average performance by 41.5\% for SpMV and SpMM, across 15 matrices and by 20\% for PageRank and BC, compared to a state-of-the-art CSR implementation \cite{tacolib}. We also compare the software-only version of SMASH against two state-of-the-art sparse matrix frameworks \cite{intel-mkl,tacolib} on a real system. We find that even with no hardware support, SMASH's bitmap encoding outperforms a state-of-the-art CSR implementation~\cite{tacolib}.

In this paper, we make the following \textbf{key contributions}:
\vspace{0.2mm}
\begin{itemize}[topsep=2pt]
    \item We show that discovering the positions of non-zero elements (\textit{indexing}) is a key bottleneck in sparse matrix computation. We demonstrate that efficient indexing can boost the performance of sparse matrix operations significantly.
    \item  We introduce SMASH, a hardware-software cooperative mechanism that enables the hardware to recognize and exploit the compression encoding used in software. SMASH consists of 1) a novel software encoding that uses a hierarchy of bitmaps to efficiently compress sparse matrices and 2) hardware support to enable highly-efficient indexing of sparse matrices that are compressed using SMASH's software encoding.  
    \item We show how SMASH can efficiently compress sparse matrices with diverse structure and sparsity characteristics using the hierarchical bitmap encoding. We design and demonstrate the effectiveness of the Bitmap Management Unit (BMU) that efficiently buffers and scans the bitmap hierarchy in hardware to identify non-zero regions in the matrix.
    We introduce an expressive ISA that enables the flexible use of SMASH in a wide variety of sparse matrix operations.
    \item  We evaluate SMASH on important sparse matrix kernels and graph analytics applications using a collection of matrices with diverse structure and sparsity. We find that SMASH provides significant performance improvements compared to state-of-the-art CSR implementations while incurring a very modest area overhead in a modern out-of-order CPU.
\end{itemize}
\vspace{-2.5mm}
\section{Motivation}
\vspace{-1mm}

\label{Backgroundlbl}


Sparse matrix operations are widely used in  a variety of applications including sparse linear algebra~\cite{sparseEIGEN,sparseLU}, graph processing~\cite{Brin1998the,besta2017slimsell}, convolutional neural networks (CNNs)~\cite{liu2015sparse}, and machine learning (ML)~\cite{dnn2018,linden2003amazon,recommenderFB2019,recommendfb2}.
These applications involve matrices with very high sparsity, i.e., a large fraction of zero elements. Using a compression scheme is a straightforward approach to avoid unnecessarily 1) storing zero elements and 2) performing computations on them. To this end, a variety of sparse matrix representation formats (e.g., ~\cite{datacompression2006,im1999optimizing,formatSurvey2016,csr52015,variableblock2005,blockoptim1999,Sengupta2007scan,bccoo2014}) have been proposed to compress the sparse matrix. The most widely used state-of-the-art format is Compressed Sparse Row (CSR)~\cite{csr52015}. In this section, we describe the CSR format and demonstrate its inefficiency.

\label{csr_description}
\vspace{-3mm}
\subsection{Compressed Storage Formats }

\begin{figure}[h!]
    \centering
     \includegraphics[width=\linewidth]{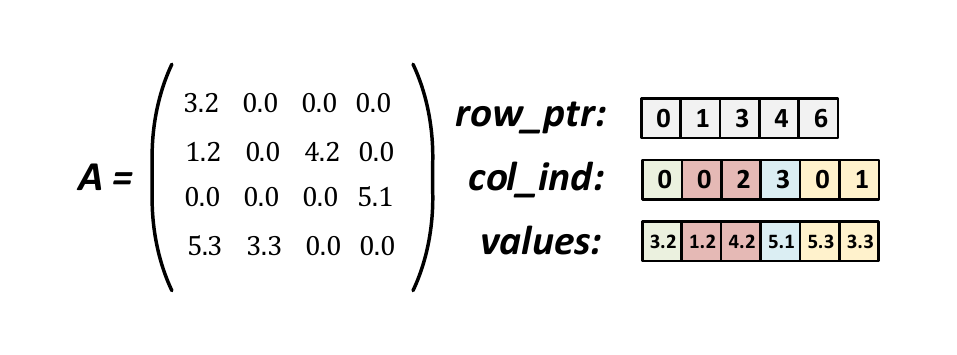}
     \vspace{-12mm}
    \caption{Compressed Sparse Row format  for a $4~\times~4$ matrix with 6 non-zero elements. We count the number of non-zero elements of row $i$ by computing \texttt{(row\_ptr[$i+1$]-row\_ptr[$i$])}. \texttt{col\_ind} holds the column index of each non-zero element.} 
    \label{fig:CSR}
    \vspace{-4mm}
\end{figure}

The Compressed Sparse Row (CSR) format ~\cite{csr52015} is widely used in many libraries that involve sparse matrix operations~\cite{intel-mkl, openblas1, openblas2, sparsex2018,tacolib}. It consists of three one-dimensional arrays: \texttt{row\_ptr}, \texttt{col\_ind}, and \texttt{values}. Given an $M~\times~N$ matrix, the \texttt{row\_ptr} array is used to store (and determine) the number of non-zero elements per row; the \texttt{col\_ind} array indicates the column indices of the non-zero elements; and the \texttt{values} array stores the values of only the non-zero elements. Discovering the position of a non-zero  element in row $i$ requires streaming through \texttt{col\_ind} from \texttt{col\_ind[row\_ptr[i]]} up to \texttt{col\_ind[row\_ptr[i+1]]} to discover its column index in the row. Figure~\ref{fig:CSR} illustrates an example of a compressed matrix using the CSR format. In this example, in order to discover the non-zero elements of row 1 (i.e., second row from the top) of A, we search in \texttt{col\_ind} starting from index \texttt{col\_ind[row\_ptr[1] == 1]} up to  \texttt{col\_ind[row\_ptr[2] == 3]}.

A variant of CSR is Compressed Sparse Column (CSC). CSC stores the elements in column-major order instead of row-major. The \texttt{col\_ptr} array is used to store (and determine) the number of non-zero elements per column, the \texttt{row\_ind} array holds the row indices of the non-zero elements, and the \texttt{values} array stores the values of the non-zero elements themselves.

CSR significantly reduces the amount of memory needed to store a sparse matrix, especially when the matrix is large and its sparsity is high. However, CSR and schemes with CSR-like structures \cite{csr52015,Sengupta2007scan} have one major requirement: in order to determine where the non-zero elements are located in the original matrix, the corresponding indices need to be retrieved from the \texttt{row\_ptr} and \texttt{col\_ind} data structures. Accessing these data structures adds many additional instructions and requires a series of indirect data-dependent memory accesses. These overheads reduce the benefits of avoiding the computation on zero elements. Hence, even though CSR-like formats reduce storage requirements and avoid needless computation, discovering the positions of the non-zero elements still is an unsolved challenge that causes performance and efficiency overheads.
\vspace{-2mm}

\subsubsection{Sparse Matrix Vector Multiplication (SpMV)}

 We consider the SpMV kernel $ y  := y + Ax$, where $A$ is a sparse matrix, $x$ is a dense vector, and $y$ is the output vector. The naive 2D implementation of the SpMV kernel involves performing computations on every element of the two-dimensional matrix A and incurs high computational and storage overheads. Code Listing \ref{code:CSR_SpMV} presents the CSR-based implementation. In this case, the algorithm iterates over only the non-zero elements and avoids unnecessary zero-element computations. However, it introduces a pointer-chasing operation when \texttt{col\_ind} is loaded and then used as an index to load the appropriate element of the vector  \texttt{(x[col\_ind[j]])}. Only after this complex indexing operation can we perform the multiplication with the corresponding non-zero element in matrix A (\texttt{values[j]}), as shown in Line 3 of Code Listing \ref{code:CSR_SpMV}.

\begin{code}
\begin{minted}[escapeinside = ||, xleftmargin=3em, linenos]{c}
 for (i = 0; i < N; i++)
    for (j = |\textcolor{blue}{\detokenize{row_ptr[i]}}|; j < |\textcolor{blue}{\detokenize{row_ptr[i+1]}}|; j++)
        y[i] += values[j] * x[|\textcolor{blue}{\detokenize{col_ind[j]}}|]
\end{minted}
\vspace{-3mm}
\captionof{listing}{CSR-based SpMV implementation. The column index of each element is needed to perform the multiplication with the x   vector.}
\label{code:CSR_SpMV}
\end{code}
\subsubsection{Sparse Matrix Matrix Multiplication (SpMM)}

\begin{figure}[ht]
    \centering
    \hspace*{-8mm}
    \includegraphics[scale =0.68]{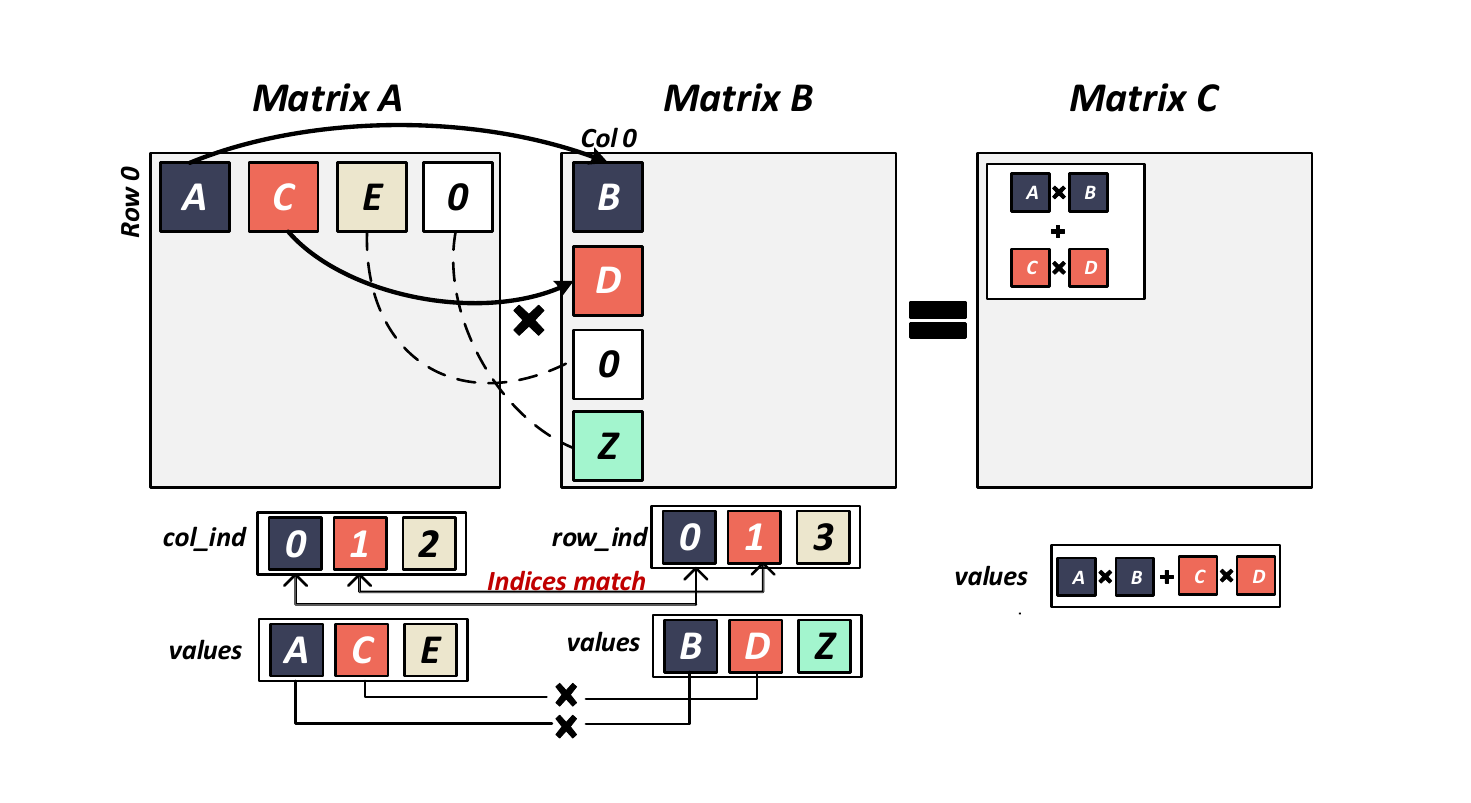}
    \vspace{-10mm}
     \caption{Index matching in SpMM. We search \texttt{col\_ind} of matrix A  and \texttt{row\_ind} of matrix B to find indices that match. Only the indices of the first two elements of A's Row 0 and B's Column 0 match. The remaining two elements in A's Row 0 or B's Column 0  have at least one zero element in them and their indices do \textbf{not} match.}
    \label{fig:SpMM_index}
    \vspace{-5mm}
\end{figure}

SpMM is traditionally performed using inner product multiplication \cite{outerspace2018}. This results in a series of dot product operations between each row of the first matrix (A) and each column of the second matrix (B) to produce the final elements of the result matrix (C). The naive $O$($n^3$) SpMM implementation  is prohibitively expensive due to the high number of unnecessary computation operations on zero elements. A CSR-based implementation of SpMM, shown in Code Listing~\ref{code:CSR-based_SpMM}, avoids such unnecessary computations. In SpMM, matrix A is compressed using CSR and matrix B using CSC. SpMM iterates over the rows of A and columns of B (Lines 1-2 in Code Listing \ref{code:CSR-based_SpMM}). For each non-zero element in each row of A, we need to search through \texttt{col\_ind} of matrix A and \texttt{row\_ind} of matrix B to discover which elements should be multiplied during each dot product. This process is called \textit{index matching} (Lines 4-6 in Code Listing \ref{code:CSR-based_SpMM}). Figure \ref{fig:SpMM_index} presents an example of index matching. We need to match the positions of the non-zero elements of matrix A at row $0$ and  the non-zero elements of matrix B at column $0$ to perform the dot product correctly. 
Given that index matching is performed for every dot product operation, a CSR-based SpMM implementation requires a large number of position-finding operations for non-zero elements and thus the indexing mechanism plays a critical role in performance and efficiency.


\begin{code}
\begin{minted}[escapeinside=||,xleftmargin=3em, linenos ]{c}
for each row of A
    for each column of B
        for each non-zero element in row of A 
            k1 = |\textcolor{blue}{\detokenize{search_in_col_ind_of_A()}}|
            k2 = |\textcolor{blue}{\detokenize{search_for_row_ind_of_B()}}|
            if (k1 == k2)
                y[i][j] += a_val[|\textcolor{blue}{\detokenize{k1}}|] * b_val[|\textcolor{blue}{\detokenize{k2}}|]
\end{minted}
\vspace{-3mm}
\captionof{listing}{CSR-based inner-product SpMM implementation. Index matching (Lines 4-6) is needed to perform the multiplication of A's rows and B's columns (line 7).}
\vspace{-7mm}
\label{code:CSR-based_SpMM}
\end{code}

\subsection{Limitations of Existing Compressed Storage Formats}
\vspace{-1mm}

Compressed storage formats, such as  CSR \cite{csr52015}, BCSR \cite{im1999optimizing}, and CSR5 \cite{csr52015}, are effective at reducing the storage area and avoiding redundant computations on the zero elements of the matrix. However, as described above, they require additional computation and indirect memory accesses to find the indices, i.e., the row and column positions, of non-zero elements. This increases the computation burden and memory traffic, and hence lowers the potential gains from the compression scheme. 

\begin{figure}[!t]
    \centering
    \includegraphics[scale = 1.0]{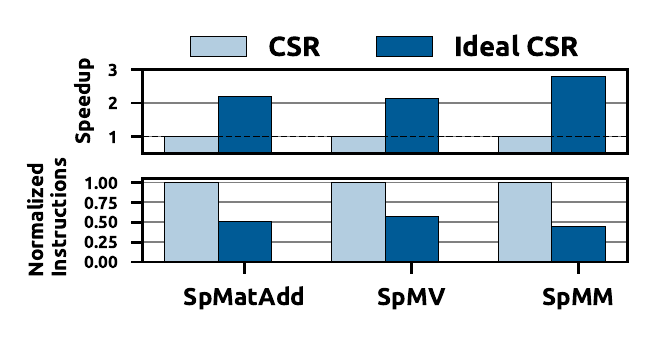}
    \vspace{-4.5mm}
     \caption{Speedup and normalized number of executed instructions of an
     ideal indexing scheme over the baseline CSR, averaged across 15 sparse matrices for Sparse Matrix Addition, SpMV, and SpMM (see Section \ref{sec:setup} for methodology).}
    \label{fig:Optimal_indexing}
    \vspace{-5.2mm}
\end{figure}

To understand the impact of this \emph{indexing overhead} on sparse matrix processing, we conduct an experiment where we compare a state-of-the-art CSR implementation to an idealized version in which accessing the positions of non-zero elements does not incur any additional computation or memory access.  Figure~\ref{fig:Optimal_indexing} shows the speedup and executed instruction count of this idealized CSR over the regular state-of-the-art CSR. As shown in the figure, eliminating the indexing part of the CSR format significantly improves performance: $2.21\times$ for Sparse Matrix Addition, $2.13\times$ for SpMV, and $2.81\times$ for SpMM. These performance benefits come from the reduced number of executed instructions (by 49\%, 42\%, 65\%, respectively) and from eliminating the expensive pointer-chasing lookups in memory.


\vspace{-2mm}
\subsection{Other Approaches to Sparse Matrix Compression }
\vspace{-0.5mm}

Other approaches ~\cite{uniquekourtis2008,csxkourtis2011,smatautotune2013,csb2009,pattern2009,cocktail2012,csradaptgpus2014} aim to maximize the efficiency of sparse matrix computations by specializing to particular matrix types and thus trading off generality. These approaches assume and leverage a specific matrix structure or known pattern. Saad et al. ~\cite{saad2003iterative} assume a diagonal matrix and base the compression scheme around the assumption that all non-zero elements are only along the matrix diagonal. Kourtis et al. ~\cite{uniquekourtis2008} assume matrices with few unique non-zero elements in designing the compression scheme. As a result, these approaches are not applicable to a wide range of important classes of applications like CNNs and graph processing algorithms, where such assumptions do not hold. 




\vspace{-1mm}
\section{SMASH : Design Overview}
We introduce SMASH, a hardware-software cooperative mechanism that 1) enables highly-compressed storage and avoids unnecessary computation; 2) significantly reduces the overheads imposed by the compression scheme itself (i.e., enables efficient discovery of the positions of non-zero elements); and 3) can be used generally across a diverse set of sparse matrices and sparse matrix operations.

The \textbf{key idea} of SMASH is to enable the hardware to recognize and exploit the compression encoding used in software. We devise a new construct, recognized by both the hardware and software, to compress sparse matrices efficiently: \emph{a hierarchy of bitmaps}. Each bitmap in the hierarchy efficiently encodes sparsity by denoting the presence/absence of non-zero values in a matrix region using a single bit. The size of the region  varies with the level of the bitmap in the hierarchy and can be adjusted by software. This representation does not assume any structure in the matrix and can be used to efficiently compress a diverse set of sparse matrices. At the same time, as we demonstrate, it enables designing hardware that can exploit the known sparsity in data and hence perform highly\kon{-}efficient indexing of sparse matrices. 
\vspace{2mm}
\subsection{Design Challenges}
\label{sec:challenges}
\vspace{-0.5mm}
Designing SMASH involves addressing two major challenges:

\textbf{Challenge 1: Efficiently Encoding Bitmaps.} Representing each element in a matrix with a single bit to denote sparsity is highly-inefficient in terms of storage and computation. Hence, we need a more efficient bitmap representation that is effective regardless of the matrix sparsity and the location/distribution of the non-zero values. At the same time, hardware should be able to flexibly interpret and leverage this bitmap encoding.

\textbf{Challenge 2: Flexibility and Expressiveness.} To express a diverse set of sparse matrix operations in any application, we need a rich cross-layer interface between the application and the underlying hardware that 1) allows the software to flexibly manipulate and index sparse matrices encoded using our hierarchical bitmap encoding and 2) enables hardware to easily interpret the sparse matrix operations in the application and effectively accelerate those operations.

\vspace{-2.5mm}
\subsection{SMASH: Key Components}
\vspace{-0.5mm}

\textbf{ \quad Hierarchy of Bitmaps\kon{.}} To address Challenge 1, we represent the positions of the non-zero values in any sparse matrix with a \emph{hierarchy} of bitmaps. Our system is designed to support a certain maximum number of levels of the hierarchy. Each level of the hierarchy encodes the presence of non-zero values with a configurable compression ratio. This compression ratio is determined by the software based on the sparsity and distribution of non-zero values in a given matrix. With this representation, a zero matrix would require only one bit and one level of the bitmap encoding. 

\begin{figure}[!h]
    \centering
    \hspace*{-9mm}
    \includegraphics[scale=1.0]{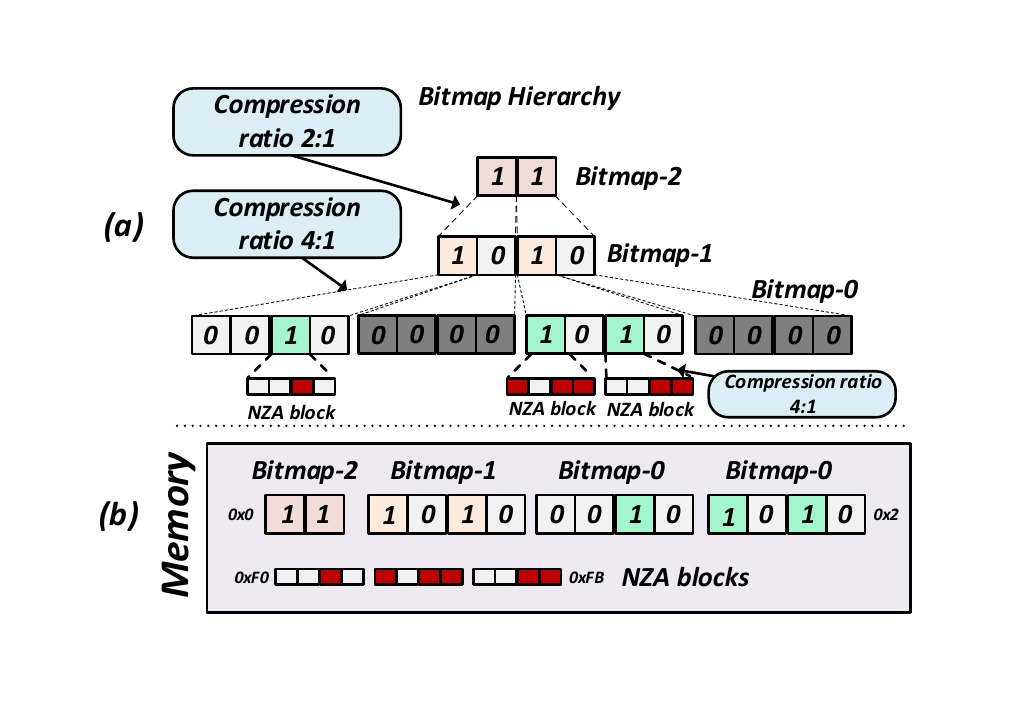}
    \vspace{-14mm}
    \caption{(a) A three-level bitmap hierarchy with different compression ratios between the levels. The NZA (Non-Zero Values Array) stores the non-zero values of the matrix. (b) We store in memory only the non-zero blocks of the bitmaps and the NZA. }
    \label{fig:overview}
    \vspace{-7mm}
\end{figure}

In Figure \ref{fig:overview}, we show an example with a 3-level bitmap hierarchy. Each bit in \emph{Bitmap-2} encodes the presence of non-zero values in two consecutive regions in \emph{Bitmap-1} (hence, the compression ratio at this level is two). \emph{Bitmap-1}, on the other hand, encodes the presence of non-zero values in \emph{four} consecutive regions in Bitmap-0 (hence, the compression ratio at this level is four). The selection of compression ratio at any level is a tradeoff between computation efficiency and storage efficiency. With a higher compression ratio, we store fewer bits to encode the presence/absence of non-zero elements but may perform unnecessary computations on zero elements. With a lower compression ratio, on the other hand, we can compute only on non-zero elements, but this would incur a higher storage overhead. 

The non-zero elements of the matrix are kept in a data structure called the  \textbf{Non-Zero Values Array (NZA)}. The granularity at which they are stored in memory depends on the compression ratio of the lowest level of the bitmap hierarchy (Bitmap-0). As we show in the example bitmap hierarchy of Figure \ref{fig:overview}, Bitmap-0 encodes the 4-element non-zero blocks of the NZA using a single bit (i.e., the compression ratio is 4:1).

Our hierarchical bitmap compression mechanism enables efficient representation of matrices with varying degrees of sparsity and varying distributions of non-zero elements by adjusting the bitmap representation granularity (i.e., compression ratios at all levels of the bitmap hierarchy). It can also be flexibly and efficiently interpreted by hardware, as we describe next.

\textbf{ Bitmap Management Unit (BMU).} We design a new hardware unit that buffers and efficiently scans the bitmap hierarchy to quickly find the non-zero elements. The BMU is responsible for efficiently and quickly calculating the indices of the non-zero regions in the matrix using the bitmap hierarchy. To this end, it caches and efficiently operates on the bitmaps in the hierarchy. Section \ref{sec:bmu} describes the BMU in detail.

\textbf{Cross-layer interface.} To flexibly accelerate a diverse range of operations on any sparse matrix, we expose to the software a rich set of primitives that 1) are general, 2) can control the BMU to quickly find the locations of non-zero elements, and 3) enable the processing core to skip unnecessary computations. These primitives are implemented as five new ISA instructions that are designed to 1) communicate  the parameters of a sparse matrix and its bitmap hierarchy to the BMU, 2) load the bitmaps into the BMU, 3) scan the bitmaps to determine the location of the non-zero elements in the matrix, and 4) communicate the <row, column> positions of the non-zero elements in the original sparse matrix back to the application. These instructions are sufficiently expressive to be used for a diverse set of sparse matrix computations, such as Sparse Matrix Vector Multiplication, Sparse Matrix Matrix Multiplication, Sparse Matrix Addition, and more \cite{sparseLU,sparseEIGEN}, thereby addressing Challenge 2 (Section \ref{sec:challenges}). Section \ref{sec:smash_isa} describes the SMASH ISA primitives in detail. Section \ref{sec:smash_use_case} shows examples of how the primitives can be used in software applications.



\vspace{-2mm}
\section{SMASH: Detailed Design}
\vspace{-0.5mm}

\label{sec:mechanism}
In this section, we provide a detailed description of the different components \kon{of} SMASH and their operation.

\vspace{-2mm}
\subsection {Software Compression (Hierarchical Bitmap Compression)}
\label{sec:hierbc}
\vspace{-1mm}

The hierarchy of bitmaps is the key construct of SMASH that enables 1) highly-compressed matrix storage and 2) efficient discovery of the positions of the non-zero regions of the matrix. The Non-Zero Values Array (NZA) holds the actual values of the sparse matrix. Every set bit of the last-level Bitmap-0 corresponds to one non-zero \emph{block} in the NZA. 
The \emph{size} of each non-zero block in the NZA (that is encoded by a single set bit in Bitmap-0) depends on the compression ratio used for Bitmap-0.

There are two major factors that impact the effectiveness of our hierarchical bitmap compression scheme: 1) the selected compression ratio at each level of the bitmap hierarchy (Section \ref{sec:trade1}) and 2) the distribution of non-zero elements in the matrix (Section \ref{sec:trade2}).

\subsubsection{Impact of compression ratio.} 
\label{sec:trade1}
\begin{figure}[!h]
    \vspace*{-9mm}
    \includegraphics[scale=1.0]{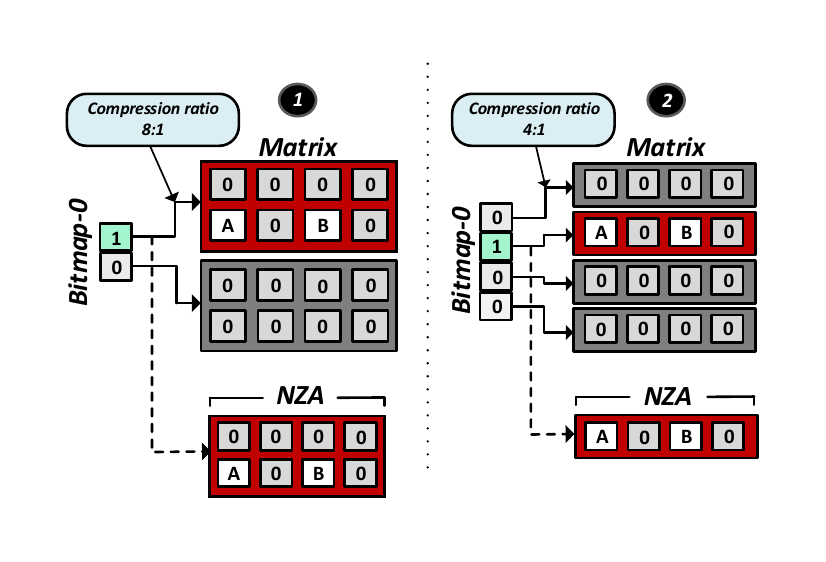}
    \vspace{-10mm}
    \caption{Two bitmaps that compress the matrix with two different compression ratios. Bitmap \protect\circled{1} encodes blocks of 8 elements with a single bit, while Bitmap \protect\circled{2} encodes blocks of 4 elements with a single bit.}
    \label{fig:Bitmap}
    \vspace*{-2.5mm}
\end{figure}

Figure \ref{fig:Bitmap} demonstrates the impact of choosing different compression ratios for Bitmap-0 using a simplified example. We show two cases where the bitmap encodes the same $4 \times 4$ matrix using two different compression ratios between Bitmap-0 and the NZA. In case \circled{1}, the bitmap uses a single bit to encode 8 consecutive elements of the matrix, i.e., the compression ratio is 8:1. The bitmap requires 2 bits to encode the entire matrix and the NZA holds one 8-element non-zero block (consisting of 2 non-zero and 6 zero elements). In case \circled{2}, the bitmap uses a \emph{single} bit to encode 4 consecutive elements of the matrix, i.e, the compression ratio is 4:1. In this case, the bitmap requires 4 bits to encode the entire matrix and the NZA holds one 4-element block (consisting of 2 non-zero and 2 zero elements). 

As we demonstrate with this example, a higher compression ratio reduces the size of the bitmap and thus scanning it becomes more efficient. However, with a higher  compression ratio, the NZA unnecessarily stores more zero elements. Since zeroes \emph{within} the block cannot be identified a priori, the processor unnecessarily performs computation on them. 

Hence, the compression ratio forms a tradeoff between 1) smaller bitmaps that can be quickly scanned and 2) more zero elements in the NZA storage and unnecessary computations on them.
This major tradeoff also applies to the higher levels of the bitmap hierarchy.

\subsubsection{Impact of distribution of non-zero elements in the sparse matrix} 
\label{sec:trade2}
The \emph{distribution} of non-zero elements across the matrix also affects the size of the NZA. On the one hand, if the non-zero elements of the matrix are closely clustered, the number of non-zero blocks of the matrix decreases and the NZA stores fewer blocks. On the other hand, if the non-zero elements of the matrix are distributed more uniformly across the matrix, the number of non-zero blocks of the matrix increases and the NZA may need to hold more non-zero blocks (that also contain more zeros).

\subsubsection{Conversion to the hierarchical bitmap format}

If the input data to any user application is already stored using another compression format (such as CSR), the application needs to convert the sparse matrices to the hierarchical bitmap encoding and the NZA used in SMASH. This is done using three steps. First, the application identifies all the non-zero blocks in the matrix using the indexing mechanism required by the original format. The size of the block depends on the assumed size of the non-zero blocks in the NZA. Second, the application creates the NZA by appending these non-zero blocks contiguously in memory. Third, the application creates the bitmap hierarchy, starting with Bitmap-0. To create Bitmap-0, the application determines the locations of the non-zero blocks of the matrix (where the block size is equal to the Bitmap 0 compression ratio).  For each one of these locations, the application sets the corresponding bit of Bitmap-0 to 1. Next, the application creates the higher levels of the bitmap hierarchy: each level $i$ of the hierarchy is created based on the compression ratio of Bitmap-$i$ and the corresponding set bits of Bitmap-$(i-1)$. Assuming the chosen compression ratio of level $i$ is 8:1, we set each bit in Bitmap-$i$ if there are one or more set bits in the corresponding 8 elements in Bitmap-$(i-1)$.

We note that this conversion process from any format to our hierarchical bitmap encoding can be automated in software, and, if needed, accelerated with hardware support.

\vspace{-1mm}
\subsection {Hardware Indexing \newline (Bitmap Management Unit)}

\label{sec:bmu}

The Bitmap Management Unit (BMU) provides the key functionality of scanning the bitmap hierarchy to quickly identify the non-zero elements of the matrix in a highly-efficient manner. It recognizes the bitmap encoding used in software based on the parameters of the bitmap (and sparse matrix) provided to it via the SMASH software-hardware interface (Section \ref{sec:smash_isa}).
\vspace{-2mm}
\subsubsection{BMU components.} Figure \ref{fig:BMU} demonstrates the structure of the BMU. It consists of four key components: 1) the SRAM buffers that hold the bitmaps when they are being scanned, 2) the hardware logic that scans the buffers to find the non-zero blocks, 3) programmable registers that hold configuration parameters, such as the matrix dimensions and the compression ratios of the bitmap hierarchy, and that effectively orchestrate the operation of the hardware logic, and 4) two registers to store the row and column indices of the non-zero elements determined by the BMU. The BMU supports multiple \emph{groups} of the components presented in Figure \ref{fig:BMU}, to enable the indexing of multiple sparse matrices, where each group is dedicated to buffering and scanning a single sparse matrix.

The SRAM buffers hold the bitmaps one block at a time. In our implementation, each buffer is 256 bytes in size. The compression ratio (the number of bytes encoded by a single bit) at each level of the bitmap, including between Bitmap-0 and the NZA, must be less than or equal to the bitmap buffer size. This is to avoid loading the buffers multiple times from memory to scan a single block, which would be expensive and inefficient. For example, with a 256-byte SRAM buffer size, the maximum compression ratio supported in the BMU is $256 \times 8$ = 2048:1.

\vspace{-1mm}
\subsubsection{BMU operation.} 
\begin{figure}[!t]
\vspace{-5mm}
\centering
\hspace*{-6mm}
\includegraphics[scale=0.81]{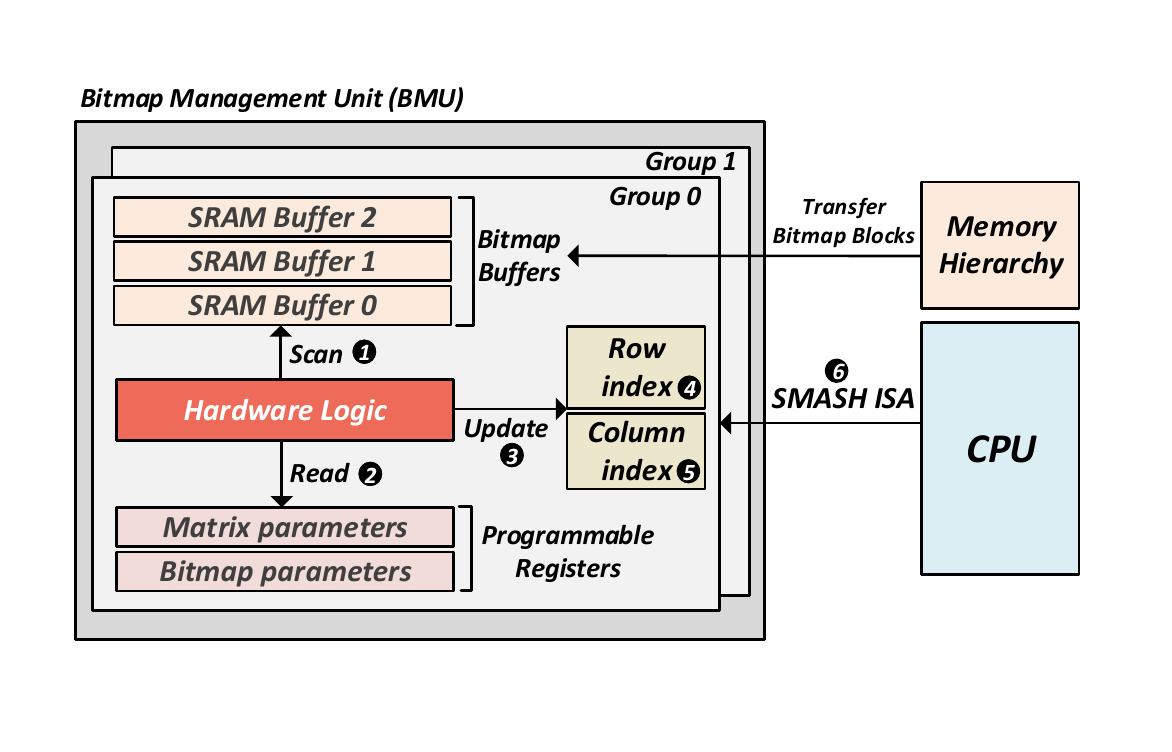}
\vspace{-14mm}
\caption{Bitmap Management Unit consists of four key components: 1) SRAM buffers to store portions of the bitmaps that are being operated on, 2) hardware logic that scans the buffers, 3) registers to hold the matrix and bitmap parameters, and 4) output registers to store the row ~\protect\circled{4} and column indices ~\protect\circled{5} of the non-zero blocks. The BMU communicates with the CPU using the SMASH ISA ~\protect\circled{6}.}
\label{fig:BMU}
\vspace{-6mm}
\end{figure}

As depicted in Figure \ref{fig:BMU}, identifying the location of a non-zero block in the matrix involves three steps.  First, the hardware logic scans the bitmap buffers to find the set bits and discover the positions of the non-zero blocks~\circled{1}. Second, the hardware logic reads the matrix dimensions and the compression ratios from the programmable registers to calculate the row and column indices of the current non-zero block~\circled{2}. Third, it updates the output registers~\circled{3} \kon{that} hold the row \circled{4} and column \circled{5} indices of the non-zero block. These registers can
be repeatedly read by the CPU to iteratively find the location of
the non-zero elements. To find the next non-zero element, the hardware logic looks for the next set bit in the bitmap block or loads the next bitmap block from the memory hierarchy to perform the search for the non-zero blocks. These operations are controlled by software with five new ISA instructions \circled{6}, which we describe in Section \ref{sec:smash_isa}.

\vspace{-1mm}

\subsubsection {Efficient indexing with the BMU}
 The BMU iteratively communicates to the CPU the row and column indices of the non-zero elements in the sparse matrix. Since we use multiple levels of indirection with the hierarchical bitmap, finding each non-zero element involves traversing the bitmap hierarchy in a depth-first manner. Every time a set bit is encountered at any bitmap level, we save that bit's index within the bitmap and then traverse the lower-level bitmap associated with that set bit. The BMU traverses the hierarchy in this manner (saving the index of the set bit at each level) until it reaches Bitmap-0. Any set bit in Bitmap-0 directly maps to a block of elements in the sparse matrix that has at least one non-zero element.  Using the saved indices of the set bits at each level of the hierarchy, as well as the corresponding compression ratios, the BMU calculates the index of the non-zero block in the original sparse matrix.  


The final index = $row\_index*matrix\_columns$ $+ column\_index$, is computed using the hierarchy of bitmaps in the following way: 

$Index = \sum_{i=0}^{levels-1}((\prod_{0}^{j=i}comp(j))*index\_bit(i))$ where $comp(j)$ is the compression ratio of Bitmap-$j$  and $index\_bit(i)$ is the index of the encountered set bit while scanning Bitmap-$i$. The row and column indices of the non-zero block in the original sparse matrix are calculated as follows: $row\_index = Index / matrix\_columns$ and $column\_index = Index \% matrix\_columns$. These indices are stored in the two output registers dedicated to the row index and the column index in the BMU. They are retrieved by the CPU using new ISA instructions (described below). The application iterates through the non-zero blocks of the NZA to compute on the non-zero values. For each consecutive non-zero block, the application reads the row and column indices from the BMU registers to identify the \kon{<row,column> location} of the non-zero block in the original sparse matrix.
\vspace{-2mm}

\subsection{SMASH ISA: Software/Hardware Interface}
\label{sec:smash_isa}
\vspace{-1mm}
We introduce five new instructions in the ISA to control the functionality provided by the BMU. Table~\ref{table:ISA} shows the instructions. These instructions are designed  to i) communicate parameters of the matrix and the bitmap hierarchy to the BMU (\texttt{MATINFO,BMAPINFO}), ii) load the bitmaps into the BMU buffers (\texttt{RDBMAP}), iii) iteratively scan the bitmaps to determine the location of the next non-zero element in the matrix (\texttt{PBMAP}), and iv) communicate the row and column indices (in the original sparse matrix) of the non-zero element back to the application (\texttt{RDIND}).

\texttt{MATINFO}: This instruction communicates the dimensions of the matrices to the BMU. It has three source operands: \texttt{row,col,} and \texttt{grp}. \texttt{row} represents the number of rows and \texttt{col} represents the number of columns of the matrix.
\texttt{grp} is used to select the group of the BMU. If the user application involves two sparse data structures, \texttt{MATINFO} needs to be executed twice (one for each matrix) before performing other operations with the BMU.

\texttt{BMAPINFO}:
This instruction communicates the compression ratio of each bitmap to the BMU.
It has three source operands: \texttt{comp} for the compression ratio, \texttt{lvl} selects the bitmap level in the hierarchy, and \texttt{grp} selects the group of the BMU.

\newcolumntype{C}[1]{>{\centering\arraybackslash}m{#1}}

\begin{table}[!h]
    \centering
    \begin{center}
          \vspace{-2mm}
      \caption{SMASH instructions}
      \vspace{-3mm}
\begin{tabular}{ | C{3.4cm} | m{3.8cm} | } 
\hline
\textbf{ISA instruction} & \textbf{Functionality}\\ 
\hhline{|=|=|}
\hline
\texttt{matinfo row,col,grp} &  Loads the matrix dimensions into the registers of the BMU. \\ 
\hline
\texttt{bmapinfo comp,lvl,grp} & Loads the compression ratio \texttt{comp} that bitmap at \texttt{lvl} operates with. \\
\hline
\texttt{rdbmap [mem],buf,grp} &  Loads bitmap starting from \texttt{[mem]} into SRAM buffer \texttt{buf}. \\
\hline
\texttt{pbmap grp} & Signals the BMU to scan the SRAM buffers and find the row and column indices of the next non-zero block. \\
\hline
\texttt{rdind rd1,rd2,grp}  & Loads into \texttt{rd1} and \texttt{rd2} the row and column indices of the current non-zero block that are stored in the  row index and column index output registers of the BMU.\\

\hline
\end{tabular}

\label{table:ISA}
\end{center}

\vspace{-3mm}
\end{table}

\texttt{RDBMAP}: This instruction loads the bitmap into the bitmap buffers in the BMU. It has three source operands: a \texttt{[mem]} location, a \texttt{buf} selector, and a \texttt{grp} selector. It loads a bitmap block starting from the address pointed by \texttt{[mem]} into the buffer \texttt{buf} of the group \texttt{grp}   .  

\texttt{PBMAP}: This instruction signals the hardware logic of the BMU to scan the bitmap buffers and find the index of the next non-zero block. The hardware logic updates the row index and column index output registers with the position of the non-zero block. This instruction has one source operand:  \texttt{grp}  selects the group of the BMU. 

\texttt{RDIND}: This instruction communicates to the CPU the row and column indices of the current non-zero block that was identified by the BMU after the execution of \texttt{PBMAP}. This essentially involves reading the {row index and column index output} registers \kon{of} the BMU. \texttt{RDIND} has two destination registers: \texttt{rd1} and \texttt{rd2}, and a \texttt{grp} selector. \texttt{RDIND} loads the row index and column index into  \texttt{rd1} and \texttt{rd2}, respectively. \texttt{grp} is used to select the group of the BMU.

\vspace{-2mm}
\subsection{An Alternative: Software-only SMASH}
\vspace{-0.5mm}

The hierarchical bitmap encoding of SMASH can be used entirely in software, as a pure software compression mechanism without any hardware support. In this case, a sparse matrix is represented using the hierarchy of bitmaps but the indexing is still performed entirely in software, i.e., the BMU and the ISA are \emph{not} used to accelerate the indexing.
If used entirely in software, one 64-byte block of the bitmap needs to be loaded using four memory load instructions. A Count Leading Zeros (e.g., CLZ in x86) bitwise instruction is needed to find the first most-significant set bit. For every set bit that is found, one bitwise AND is needed to mask the set bit and then search for the next one. This adds more computation to find the set bits compared to the mechanism we describe in Section \ref{sec:bmu}. In contrast to Software-only SMASH, the BMU loads the whole block at once and does not need AND operations to find set bits. As we show in Section \ref{sec:eval_smk}, Software-only SMASH  cannot leverage the full benefits of the hierarchical bitmap encoding and incurs additional computational overhead.  Even so, as we also show in Section \ref{sec:eval_smk}, Software-only SMASH still outperforms CSR on average, because it uses fewer instructions overall.


\vspace{-1mm}
\section{SMASH Example Use Cases}
\label{sec:smash_use_case}
\vspace{-0.5mm}
We describe in detail how the SpMV and SpMM operations are performed using SMASH. We assume a 3-level bitmap hierarchy for SpMV and, for simplicity of explanation, a 1-level bitmap for SpMM. Our SpMM example consists of two sparse data structures.
\vspace{-1mm}
\subsection{Example Use Case 1: SpMV}
\begin{figure}[!h]
\vspace{-3mm}
\centering
 \includegraphics[width=\linewidth]{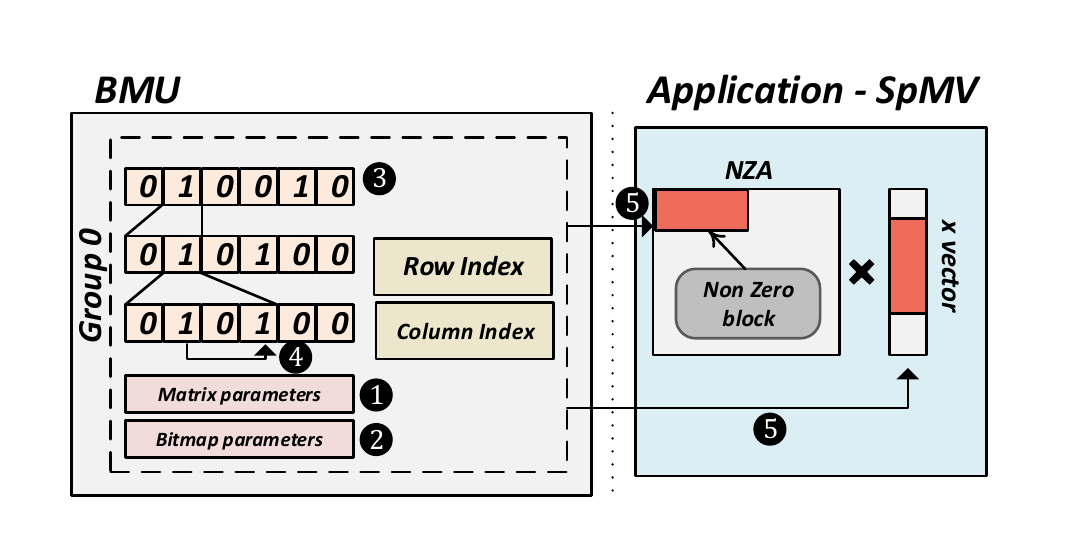}
 \vspace{-10mm}
 \caption{SpMV flow of execution. One matrix is compressed using a 3-Level Bitmap Hierarchy. }
 \label{fig:Flow-SpMV}
\vspace{-2.5mm}
 \end{figure}

\vspace{-0.5mm}
Figure \ref{fig:Flow-SpMV} and Algorithm  \ref{alg:SpMV_OP} describe the execution flow for SpMV using SMASH. SpMV involves only one sparse matrix and a dense vector, x. Therefore, it utilizes only one group of the BMU's components. \texttt{MATINFO} is used in the beginning of the algorithm to communicate the dimensions of the matrices \circled{1} to the BMU (line 2 in Algorithm \ref{alg:SpMV_OP}). \texttt{BMAPINFO} is executed once for each bitmap to communicate the compression ratios \circled{2} (lines 3-5).  \texttt{RDBMAP} is executed three times at the beginning to load the bitmap hierarchy into the bitmap buffers~\circled{3} (lines 6-8). \ Whenever a non-zero block is found, \texttt{PBMAP} is used to search in the SRAM buffers to find the row and column indices of the next non-zero block of the NZA~\circled{4} (line 11) and
\texttt{RDIND} returns these indices of the non-zero block back to the application (line 12). The processor loads the block from the NZA,  and multiplies it with the x vector's block at the row index returned by \texttt{RDIND} \circled{5} (lines 15-16).

     

\setlength{\textfloatsep}{2pt}
\begin{algorithm}[!h]
\caption{: SpMV using SMASH}
\begin{lstlisting}[deletendkeywords={filter}]
# Operation: A * x = C
matinfo rows,columns,0 # Load dimensions to BMU
bmapinfo comp2,2,0
bmapinfo comp1,1,0 # Load compression ratios to BMU
bmapinfo comp0,0,0
rdbmap [bitmap2],2,0
rdbmap [bitmap1],1,0 # Load 3 bitmaps in BMU buffers
rdbmap [bitmap0],0,0
ctrNZ = 0 # Initialize counter of NZ blocks
for all non-zero blocks of the sparse matrix 
    pbmap 0 # Scan the bitmaps
    rdind rowInd,colInd,0 # Read index of the NZ block
    ctrElmt = 0 # Initialize counter of elements
    for all elements of block (rowInd,colInd) 
         NZA_ind = ctrNZ*comp0 + ctrElmt
         C[rowInd+ctrElmt]+=NZA[NZA_ind]*x[colInd+ctrElmt]
         ctrElmt+=1 # Point to the next element
    ctrNZ+=1 # Point to the next NZ block
\end{lstlisting}
\label{alg:SpMV_OP}
 \end{algorithm}


\subsection{Example Use Case 2: SpMM}
\begin{figure}[!h]
  \vspace{-8mm}
\centering
 \includegraphics[width=\linewidth]{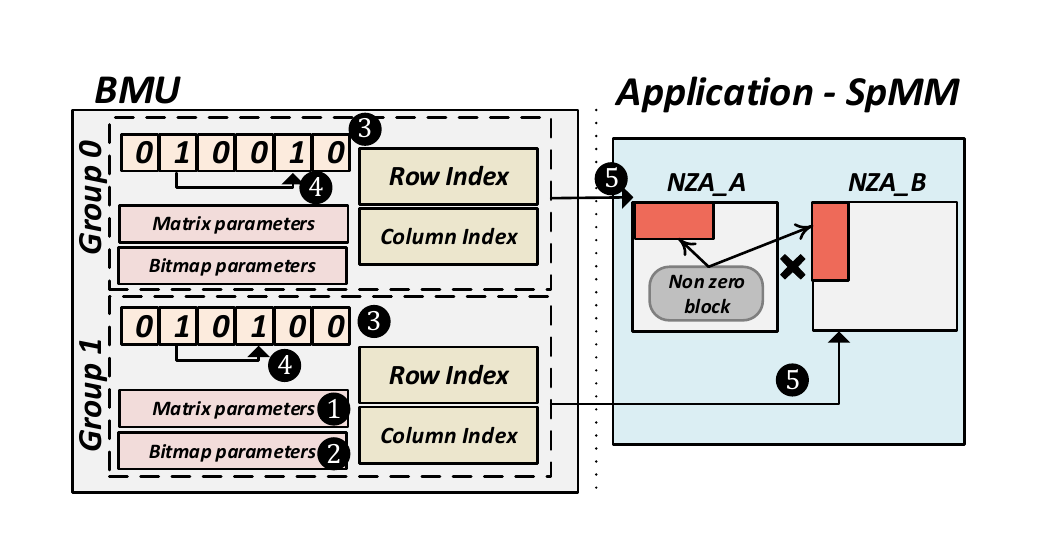}
 \vspace{-10mm}
 \caption{SpMM flow of execution. Two  matrices are compressed, each using a single-level Bitmap Hierarchy. }
 \label{fig:Flow-SpMM}
\vspace{-4.5mm}
 \end{figure}

\vspace{-0.5mm}
Figure \ref{fig:Flow-SpMM} and Algorithm \ref{alg:SpMM_OP} describe the execution flow for SpMM using SMASH. 
    We describe SpMM in the case where a 1-level bitmap hierarchy is used for each of the two sparse data structures.
    In the initialization phase of the algorithm, we need to execute \texttt{MATINFO} twice, one for each sparse matrix \circled{1} (lines 2-3 in Algorithm \ref{alg:SpMM_OP}). We also need to execute \texttt{BMAPINFO} twice, one for each bitmap~\circled{2} (lines 4-5).
    The program iterates over the rows of matrix A and columns of matrix B. For each row of matrix A, we load the bitmap at the correct offset using \texttt{RDBMAP} as follows: \texttt{[mem]=}\textit{bitmapA+rowOffset}, where \texttt{buf=0} and \texttt{grp=0} \circled{3} (line 7). For each column of matrix B, we load the bitmap at the correct offset in the second group of buffers using \texttt{RDBMAP} as follows: \texttt{[mem]=}\textit{ bitmapB+colOffset}, where \texttt{buf=0} and \texttt{grp=1} (line 9). Index matching requires the use of the \texttt{PBMAP} instructions~\circled{4} (lines 10-11) to search the bitmaps of both matrices and determine the matching indices of NZA\_A's and NZA\_B's blocks. \kon{\texttt{RDIND} instructions~\circled{5} (lines 12-13) are executed to load the indices of the non-zero blocks into registers so that the column index of A can be compared to the row index of B (line 14). \kon{If the indices match, the algorithm performs the inner product between the corresponding blocks of NZA\_A and NZA\_B (line 15).} \kon{If the row index of A is greater than the current row or if the column index of B is greater than the current column of B (line 16), the algorithm skips the remaining inner-product computation (lines 10-15) between the current row of A and the current column of B. This implies the absence of anymore non-zero elements in the current row of A (or column of B) and thus, unnecessary computation on zero elements is avoided.}}

The \kon{index matching phase of the algorithm requires calculating the indices of each non-zero element in both A and B for each inner product (line 10-16). A format like CSR would incur extra computations and expensive indirect memory accesses to perform this step \kon{(i.e., the index matching)}. With SMASH, we leverage the BMU to accelerate the index matching. We demonstrate the benefits of our scheme in Section \ref{sec:evaluation}}.

\vspace{-2mm}

\setstretch{0.97}
\subsubsection {Generality of SMASH for other use cases}
In this work, we evaluate SMASH using two sparse matrix kernels, SpMV and SpMM, that are central to many important applications (e.g., ~\cite{van2008graph,liu2015sparse,penn2006efficient}). However, SMASH is generally applicable to \emph{any} sparse matrix computation. Sparse matrix computations operate only on the non-zero elements of the matrix and hence fundamentally require 1) identifying the indices (row and column) of the non-zero elements and 2) retrieving those non-zero elements from memory. With the proposed ISA instructions, we can efficiently determine the location (in the matrix) of the non-zero blocks of the NZA, regardless of 1) the computation that will be performed on the non-zero elements and 2) the structure and the extent of sparsity in the matrix.
Other examples of widely used operations on sparse matrices that SMASH can accelerate include Sparse LU Decomposition~\cite{sparseLU}, Sparse Eigenvalue Calculation~\cite{sparseEIGEN,sparse-EIGEN2} and Sparse Iterative Solvers ~\cite{saad2003iterative}.
\setstretch{0.99}

\begin{algorithm}[!h]
\caption{: SpMM using SMASH}
\begin{lstlisting}[deletendkeywords={filter}]
# Operation: A * B = C 
matinfo rowsA,colsA,0 # Load dimensions to BMU
matinfo rowsB,colsB,1
bmapinfo comp0_A,0,0  # Load compression ratios to BMU
bmapinfo comp0_B,0,1
for i in [0 .. rowsA)  # Iterate over rows of A
    rdbmap [bitmapA+rowOffset],0,0 # Load bitmap_A
    for j in [0 .. colsB) # Iterate over columns of B
        rdbmap [bitmapB+colOffset],0,1  # Load bitmap_B
        do:  pbmap 0 # Find next non-zero block
             pbmap 1 # in both matrices
             rdind rowIndA,colIndA,0 # Read NZ index in A
             rdind rowIndB,colIndB,1 # Read NZ index in B
             if (colIndA  == rowIndB) # Index matching
                C[rowIndA][colIndB]+=inner_pr(NZA_A,NZA_B)
        while (rowIndA < i) && (colIndB < j)
 
\end{lstlisting}
\label{alg:SpMM_OP}
 \end{algorithm}
 
 \vspace{-2mm}



\vspace{-1mm}
\section{Experimental Setup}
\label{sec:setup}
\vspace{-0.5mm}
We model and evaluate SMASH using the zsim simulator ~\cite{sanchez2013zsim}. Table \ref{table:Zsim_parameters} provides the system configuration we evaluate.  We simulate each workload until completion. We use an Intel Xeon system \cite{intelxeon} to perform experiments on real hardware. Table \ref{table:xeon} provides the  configuration of the real system.

\begin{table}[h]
    \centering
    \caption{Simulated system configuration}
    \label{table:Zsim_parameters}
    \vspace{-4mm}
    \footnotesize
    \resizebox{\columnwidth}{!}{%
    \begin{tabular}{@{} l @{}}
    \toprule
    \textbf{CPU} 3.6~GHz, Westmere-like \cite{westmere} OOO, 4-wide issue; \\   128-entry ROB; 32-entry LQ and SQ;  \\        
    \midrule
	\textbf{L1 Data + Inst. Cache} 32~KB, 8-way, 2-cycle; 64~B line; LRU policy;
	\\ MSHR size: 10; Stride prefetcher; \\
    \midrule
    \textbf{L2 Cache}  256~KB, 8-way, 8-cycle; 64~B line; LRU policy; \\
        MSHR size: 20; Stride prefetcher; \\
    \midrule
    \textbf{L3 Cache} 1MB , 16-way, 20-cycle; 64~B line; LRU policy; \\ 
    MSHR size: 64; Stride prefetcher; \\
    \midrule
    \textbf{DRAM} 1-channel; 16-bank; open-row policy;
    4GB DDR4;\\
    \midrule
    \end{tabular}
    }
 \vspace{-2mm}
\end{table}

\begin{table}[h]
\centering
\small
\begin{small}
 \vspace{-1mm}
\caption{Evaluated sparse matrices}
 \vspace{-3mm}
 \begin{tabular} {l l  r r c}
 \toprule
 \vspace{-1.5mm}
& \thead{Name} & \hspace{-03mm}\# \thead{Rows} & \thead{Non-Zero \\ Elements} & \hspace{-01mm}\thead{Sparsity (\%)} \\ [0.4ex] 
\midrule

M1:  & descriptor\_xingo6u	& \hspace{-03mm}20,738	 	 & 73,916	 & 0.01\\
M2:  & g7jac060sc	& \hspace{-03mm}17,730 &  183,325	 & 0.06\\
M3:  & Trefethen\_20000	& \hspace{-03mm}20,000	& 554,466	 & 0.14\\
M4:  & IG5-16	& \hspace{-03mm}18,846	 & 588,326	 & 0.17\\
M5:  & TSOPF\_RS\_b162\_c3	& \hspace{-03mm}15,374	 & 610,299	 & 0.26\\
M6:  & ns3Da	& \hspace{-03mm}20,414	& 1,679,599	 & 0.40\\
M7:  & tsyl201	&\hspace{-03mm} 20,685	& 2,454,957	 & 0.57\\
M8:  & pkustk07	& \hspace{-03mm}16,860	 & 2,418,804	 & 0.85\\
M9:  & ramage02	& \hspace{-03mm}16,830	 & 2,866,352	 & 1.01\\
M10:  & pattern1	& \hspace{-03mm}19,242	 & 9,323,432	 & 2.52\\
M11:  & gupta3	& \hspace{-03mm}16,783	 & 9,323,427	 & 3.31\\
M12:  & nd3k	& \hspace{-03mm}9,000	& 3,279,690	 & 4.05\\
M13:  & human\_gene1	&\hspace{-03mm} 22,283	 & 24,669,643	 & 4.97\\
M14:  & exdata\_1	& \hspace{-03mm}6,001 & 2,269,500	 & 6.30\\
M15:  & human\_gene2	& \hspace{-03mm}14,340	 & 18,068,388	 & 8.79\\

 \bottomrule
 \end{tabular}
 \label{table:Sparse_matrices_collection}
 \end{small}
 \vspace{-1mm}
\end{table}

\begin{table}[h]
\begin{minipage}[t]{0.25\textwidth}

\centering
\footnotesize
\caption{Input graphs}
\tabcolsep=0.05cm
 \vspace{-3mm}
 \begin{tabular} {l l l l l }
 \toprule
  \vspace{-1mm}
& \thead{Graph} & \hspace{-03mm} \thead{Vertices} & \thead{Edges}  \\ 
\midrule

G1:  & com-Youtube	& 1.1M & 2.9M	  \\
G2:  & com-DBLP	&  317K 	 & 1M	 & \\
G3:  & roadNet-CA	&1.9M & 2.7M	 \\
G4:  & amazon0601	& 403K & 3.3M	  \\

 \bottomrule
 \end{tabular}
 \label{table:graphs}
 \vspace{-2mm}
\end{minipage}%
\begin{minipage}[t]{0.2\textwidth}
\centering
\caption{Real system \\ configuration}
    \label{table:xeon}
\vspace{-4mm}
     \footnotesize
     \hspace*{-7mm}
    \begin{tabular}{@{} l @{}}
    \toprule
    \textbf{CPU} Intel Xeon Gold 5118 \\2.30 GHz 14nm \cite{intelxeon} \\        
    \midrule
    \textbf{L1} 384 KB, 8-way\\
    \midrule
    \textbf{L2} 12 MB, 16-way \\
    \midrule
    \textbf{L3} 16.5MB, 11-way \\
    \midrule
    \textbf{Main memory } DDR4-2400\\
    \midrule
    \end{tabular}
    
\end{minipage}
\vspace{-1mm}
\end{table}

\textbf{Workloads: Sparse Matrix Kernels.} We evaluate SMASH using two sparse matrix kernels, Sparse Matrix Vector Multiplication and Sparse Matrix Matrix Multiplication. We use the TACO library's \cite{tacolib} respective implementations of these kernels as our baseline. For input datasets, we use a diverse set of 15 sparse matrices from the Sparse Suite Collection \cite{florida}. The matrices have different sparsities and distributions of non-zero elements. The term sparsity refers to the fraction of non-zero elements in the matrix over the total number of elements. Table \ref{table:Sparse_matrices_collection} presents these matrices, sorted in ascending order of their sparsity. We use the term $M_{i}$ to refer to each matrix in Section~\ref{sec:evaluation}. We open source SMASH's software implementations of these sparse matrix kernels \cite{smash}.

 \textbf{Workloads: Graph Processing.} We implement PageRank and Betweenness Centrality from the Ligra Benchmark Suite ~\cite{pushpull} as SpMV computation and evaluate the performance of SMASH over the default CSR-based versions of these two algorithms. 
PageRank \cite{pagerank1999} was first used by Google to rank website pages. Specifically, it takes as input a graph and computes the rank of each vertex, which represents the relative importance of each node (e.g., webpage). PageRank iteratively uses SpMV to calculate the ranks of nodes in the graph \cite{prspmv2010}. 

 Betweenness Centrality \cite{bc1977} is a measure of the significance of each vertex based on the number of shortest paths that pass through it. Betweenness Centrality  iteratively uses SpMV to perform breadth-first searches in the graph  \cite{pushpull}. We evaluate these two workloads using a set of four graph inputs from the Sparse Suite Collection \cite{florida}. We use the term $G_{i}$ to refer to each graph input in Section \ref{sec:evaluation}.

\vspace{-2mm}

\section{Evaluation Results}
\label{sec:evaluation}

We evaluate five different mechanisms: 
\textit{(i)} \textbf{TACO-CSR}:  The CSR-based implementation from the TACO library \cite{tacolib}; \textit{(ii)}
\textbf{TACO-BCSR}: The BCSR-based implementation from the TACO library \cite{tacolib}; 
\textit{(iii)} \textbf{MKL-CSR}: CSR-based SpMV and SpMM implementations from Intel MKL \cite{intel-mkl}; \textit{(iv)}
\textbf{Software-only SMASH}: SMASH's hierarchical bitmap encoding implemented purely in software without the BMU; and \textit{(iv)} \textbf{SMASH}: our
complete proposed scheme,  with the hierarchical bitmap encoding and the BMU. 
The matrices we evaluate vary in
sparsity and hence, for SMASH implementation of each matrix,  we use different compression ratios in the bitmap hierarchy for SMASH . We denote the bitmap
configuration of each matrix (and graph) $i$ as $M_{i}.b2.b1.b0$, where $M_{i}$ denotes the matrix id and $b2.b1.b0$
denotes the compression ratios of each level in the bitmap hierarchy. We evaluate 4 different use cases: SpMV, SpMM
(Section~\ref{sec:eval_smk}), PageRank, and Betweenness Centrality (Section~\ref{sec:eval_gr}).  

\vspace{-2mm}

\subsection{Software-only Approaches}
\vspace{-1mm}
We first compare the performance of three state-of-the-art sparse matrix formats (TACO-CSR~\cite{tacolib}, TACO-BCSR~\cite{tacolib}, and
MKL~\cite{intel-mkl}) and the hierarchical bitmap encoding used in SMASH (i.e.,
Software-only SMASH) on our real Intel Xeon system (Table \ref{table:xeon}). Our
goals with this experiment are to 1) compare existing state-of-the-art software solutions to identify a baseline for our simulation experiments
and 2) evaluate the performance of SMASH's hierarchical bitmap encoding \emph{without} any hardware support. The TACO and MKL formats employ
a range of software optimizations that are orthogonal to the matrix format itself and the indexing mechanism. To ensure a fair
comparison, our implementation of Software-only SMASH includes all the software optimizations used by the TACO compiler, but
uses SMASH's hierarchical bitmap encoding instead of CSR.

\begin{figure}[h]
    \centering
    \vspace{-3.8mm}
    \includegraphics[scale = 0.8]{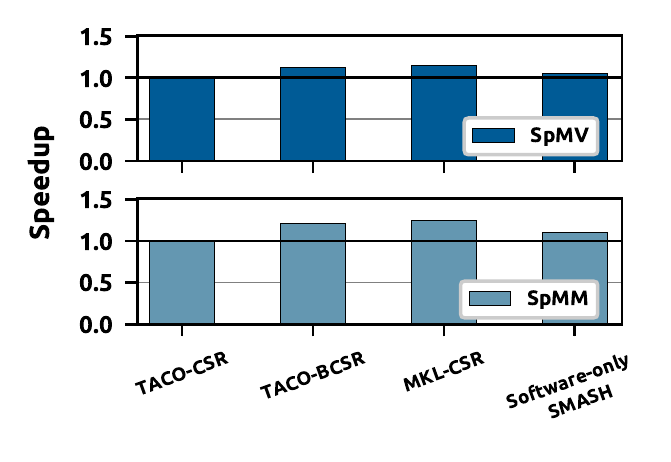}
    \vspace{-6.5mm}
    \caption{Performance of software-only approaches on a real
	Intel Xeon system (normalized to TACO-CSR).}
    \label{fig:soft}
\end{figure}

Figure \ref{fig:soft} depicts the speedup of TACO-BCSR, MKL, and software-only SMASH, normalized to TACO-CSR, for SpMV and SpMM.
We make two observations, averaged across the 15 matrices we evaluate.
First, MKL outperforms TACO-CSR in both SpMV and SpMM by 15\% and 25\% respectively. MKL also
outperforms TACO-BCSR, but by a smaller margin: 3\% in SpMV and 4\% in SpMM. While MKL uses the same CSR format as the
TACO compiler, it also employs a range of  proprietary software optimizations that lead to better performance than the TACO
implementations. We can only compare to the TACO implementations in the simulation experiments as the additional optimizations in
the closed-source MKL library cannot be added to SMASH (or to any other technique) to enable a fair and insightful comparison. 
\setlength{\textfloatsep}{2pt}
Second, we observe that Software-only SMASH outperforms TACO-CSR by 5\% in SpMV and 10\% in SpMM, but is
outperformed by both TACO-BCSR and MKL. Software-only SMASH incurs a performance overhead because 1) indexing the hierarchical bitmap
\emph{entirely} in \emph{software} requires more instructions than indexing the CSR format and 2) unlike CSR, which eliminates all zeros,  SMASH's hierarchical bitmap encoding may require unnecessary computations on zero values, depending on the choice of compression ratios in the bitmap hierarchy. 
However, despite these overheads, the hierarchical bitmap encoding avoids the expensive indexing and pointer-chasing operations in
CSR and outperforms TACO-CSR.


\vspace{-2mm}
\subsection{Sparse Matrix Kernels}
\label{sec:eval_smk}
\vspace{-0.5mm}

\subsubsection{Performance Results}
\begin{figure}[!h]
    \centering
    \hspace{-2.5mm}
    \includegraphics[scale = 0.78]{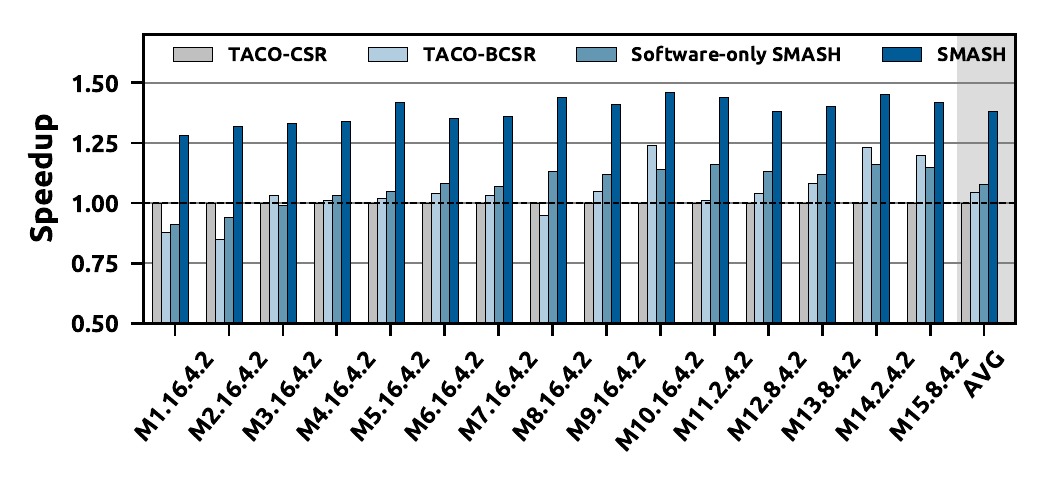}
    \vspace{-6mm}
   \caption{Speedup (normalized to TACO-CSR) for SpMV.}
    \label{fig:speedup_spmv}
        \vspace{-5.5mm}
\end{figure}

\begin{figure}[!h]
    \centering
    \hspace{-2.5mm}
    \includegraphics[scale = 0.78]{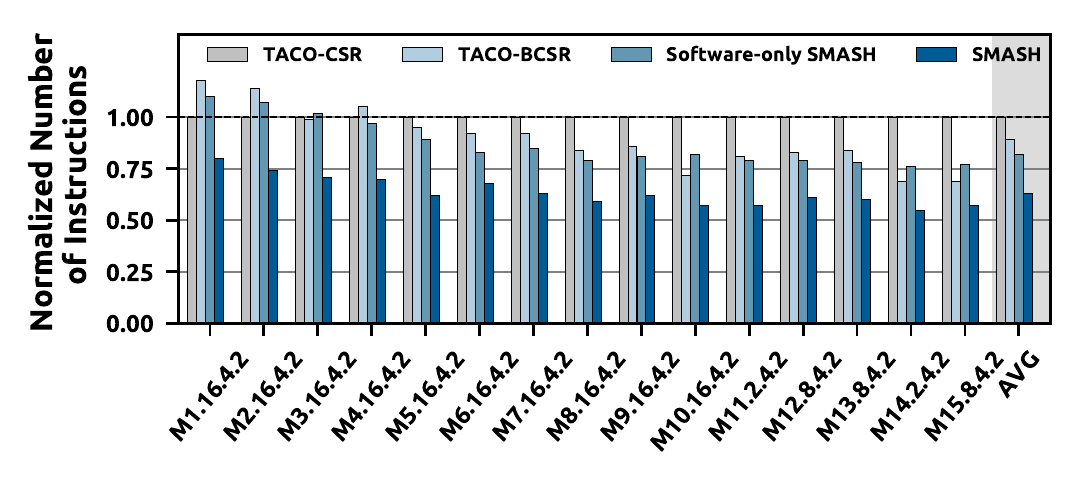}
    \vspace{-8.5mm}
   \caption{Number of executed instructions (normalized to TACO-CSR) for SpMV.}
    \label{fig:instructions_spmv}
        \vspace{-2.5mm}
\end{figure}

Figure \ref{fig:speedup_spmv} shows the speedup of TACO-CSR, TACO-BCSR, Software-only SMASH, and SMASH,
normalized to TACO-CSR, for the SpMV kernel.
Figure \ref{fig:instructions_spmv} shows the number of executed instructions in each mechanism, normalized to TACO-CSR. We make three observations.
\vspace{-2.5mm}
\vspace{-2mm}

 First, SMASH significantly outperforms all other mechanisms: it is 38\% faster over TACO-CSR and 32\% over TACO-BCSR, on average. SMASH's speedup is mainly due to executing fewer indexing instructions (47\% less than TACO-CSR
and 30\% less than TACO-BCSR, on average) and avoiding expensive pointer-chasing operations in memory. Second, SMASH is
highly effective regardless of the sparsity of the matrix. For the matrices with the highest sparsity in our evaluation
($M_{1}$ and $M_{2}$), TACO-BCSR is inefficient because it encodes data in blocks, which leads to unnecessary computation on zero elements. 
SMASH avoids such overhead by leveraging the configurability of compression ratios in our hierarchical bitmap encoding to adapt the compression ratio to the matrix
sparsity. Third, we observe that the effectiveness of Software-only SMASH depends on the \emph{sparsity} of the matrix.
Software-only SMASH incurs a higher performance overhead when the sparsity is higher because Software-only SMASH performs more unnecessary
computation on zero elements and executes more instructions searching the bitmaps for non-zero bits. In these cases,  TACO-CSR outperforms Software-only SMASH. When the
matrix is denser, the benefits of avoiding pointer-chasing and indirect indexing outweigh the additional computation
and search operations on zero elements. Thus, Software-only SMASH outperforms both TACO-CSR and TACO-BCSR for less sparse matrices. This strong impact of sparsity on overall performance is not seen as strongly
in SMASH because the hardware support makes
the indexing highly efficient even when the matrix is extremely sparse. As a result, SMASH outperforms all other approaches regardless of the matrix sparsity.

\begin{figure}[h]
    \centering
    \hspace{-2.5mm}
    \includegraphics[scale = 0.8]{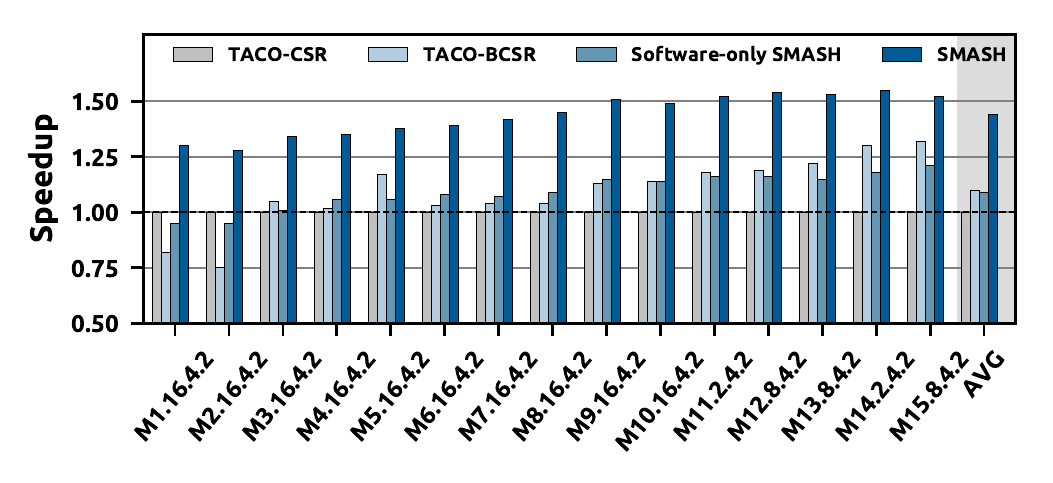}
    \vspace{-5.5mm}
   \caption{Speedup (normalized to TACO-CSR) for SpMM.}
    \label{fig:speedup_spmm}
        \vspace{-6.3mm}
\end{figure}

\begin{figure}[h]
    \centering
    \hspace{-2.5mm}
    \includegraphics[scale = 0.78]{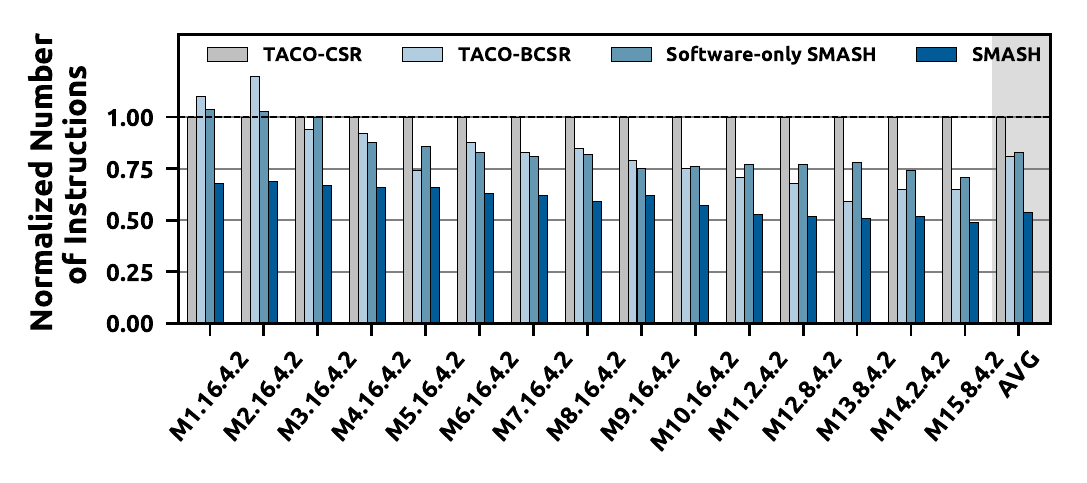}
    \vspace{-8.5mm}
   \caption{Number of executed instructions (normalized to TACO-CSR) for SpMM.}
    \label{fig:instructions_spmm}
        \vspace{-1.5mm}
\end{figure}

Figure\kon{s} \ref{fig:speedup_spmm} and \ref{fig:instructions_spmm} depict the speedup and the number of executed instructions, respectively, for the SpMM kernel.
\vspace{-4.5mm}

\vspace{-2.5mm}


We observe the same trends and make the same observations for SpMM as we do for SpMV, but with an important difference. SpMM requires \emph{twice} the number of
indexing operations as SpMV per dot product. As a result, the performance benefits of SMASH are even higher over TACO-CSR for SpMM than for
SpMV: SMASH outperforms TACO-CSR by 44\% and TACO-BCSR by 30\%, on average.

We conclude that SMASH is a highly effective mechanism to accelerate sparse matrix computation, regardless of the
sparsity of the matrix, and it significantly outperforms state-of-the-art compression
schemes for all matrices we evaluate.

\vspace{-1mm}

\subsubsection{\kon{Sensitivity to Compression Ratio}}
\label{spmvGranularitylbl}

\begin{figure}[h]
    \centering
    \hspace{-3.5mm}
    \includegraphics[scale = 0.76]{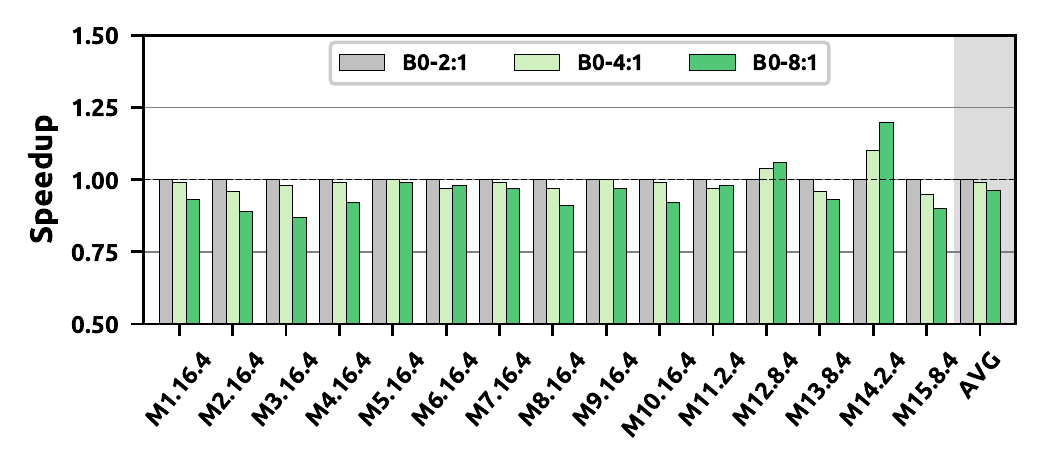}
    \vspace{-5.5mm}
   \caption{Sensitivity of SMASH speedup to the compression ratio between Bitmap-0 and the NZA for
   SpMV.}
    \label{fig:granularity-SpMV}
    \vspace{-8mm}
\end{figure}

\begin{figure}[h]
    \centering
    \hspace{-3.5mm}
    \includegraphics[scale = 0.76]{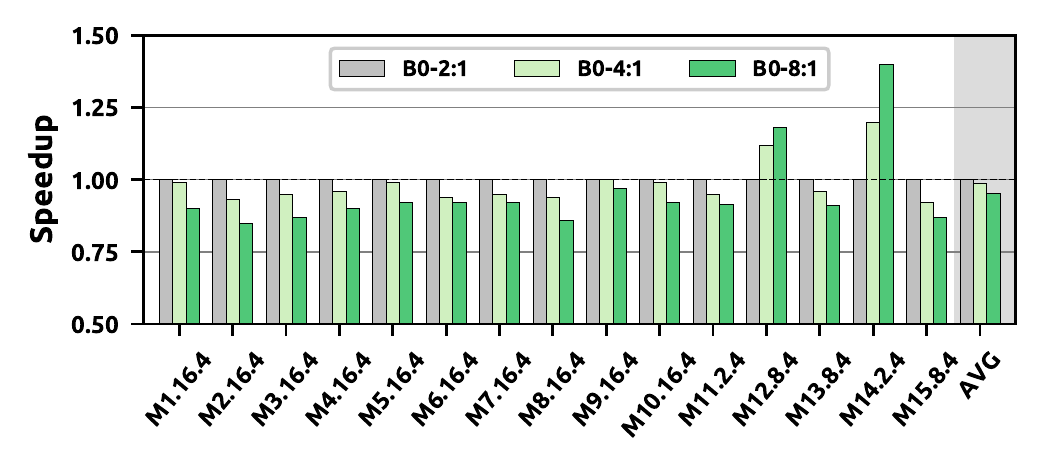}
    \vspace{-4.5mm}
   \caption{Sensitivity of SMASH speedup to the compression ratio between Bitmap-0 and the NZA for SpMM.}
    \label{fig:granularity-SpMM}
    \vspace{-3mm}
\end{figure}

As discussed in Section \ref{sec:trade1}, the compression ratio between Bitmap-0 and the NZA
defines an important tradeoff between 1) smaller bitmaps and 2)
more zero elements in the NZA that lead to unnecessary computation on zero
elements.  In Figures~\ref{fig:granularity-SpMV} and \ref{fig:granularity-SpMM},
we quantitatively evaluate this tradeoff for SpMV and SpMM respectively, by showing the speedups SMASH provides with different compression ratios. The
speedups are normalized to a configuration that uses a 2:1
compression ratio between Bitmap-0 and the NZA (i.e., each bit in Bitmap-0
encodes two elements in the NZA). 

We make two observations. First, increasing the compression ratio from 2:1 to
8:1 degrades performance, by 4\% on average (up to 13\%) for SpMV and
by 5\% on average (up to 15\%) for SpMM. This is a direct result of more
unnecessary
computations on zero elements in the NZA which cannot be skipped as a result
of the high compression ratio.\footnote{Note that we assume there is enough memory to store matrices with any bitmap compression ratio.} Second, we observe that matrices with higher
density can in some cases benefit from a higher compression ratio
(performance increases by 18\% in $M_{12}$ and 40\% in $M_{14}$, by going from a compression ratio of 2:1 to 8:1). These matrices exhibit a more
\emph{clustered} distribution of
non-zero values (i.e., the non-zero elements are close to each other). As a
result, even though the compression ratio is higher, the number of zeros in the
NZA (and hence the unnecessary computation on zero elements) does not increase
in proportion. Instead, SMASH benefits from
scanning smaller bitmaps during the indexing operation with a higher compression ratio.
 
\vspace{-4.5mm}

We conclude that the compression ratio between Bitmap-0 and the NZA plays an
important role on the effectiveness of SMASH. Our evaluations indicate that a
compression ratio of 2:1 is most effective on average and should be used
when the structure of sparsity in unknown. However, a matrix that has a \emph{known}
clustered distribution of non-zero elements may significantly benefit from a
higher compression ratio.



\vspace{-1mm}
\subsubsection{Locality of Sparsity}
\label{sec:eval_locality}
\begin{figure}[!h]
    \centering
    \hspace{-3.5mm}
    \includegraphics[scale = 0.65]{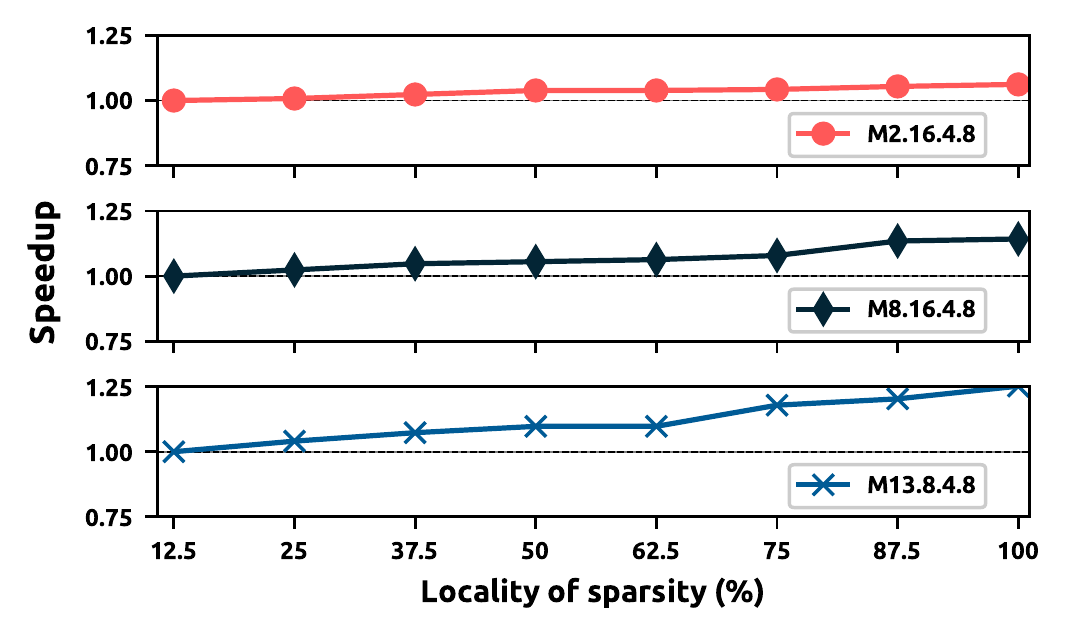}
    \vspace{-4.5mm}
   \caption{Sensitivity to locality of sparsity in SpMV.}
    \vspace{-8mm}
    \label{fig:locality-spmv}
\end{figure}

\begin{figure}[!h]
    \centering
    \hspace{-3.5mm}
    \includegraphics[scale = 0.65]{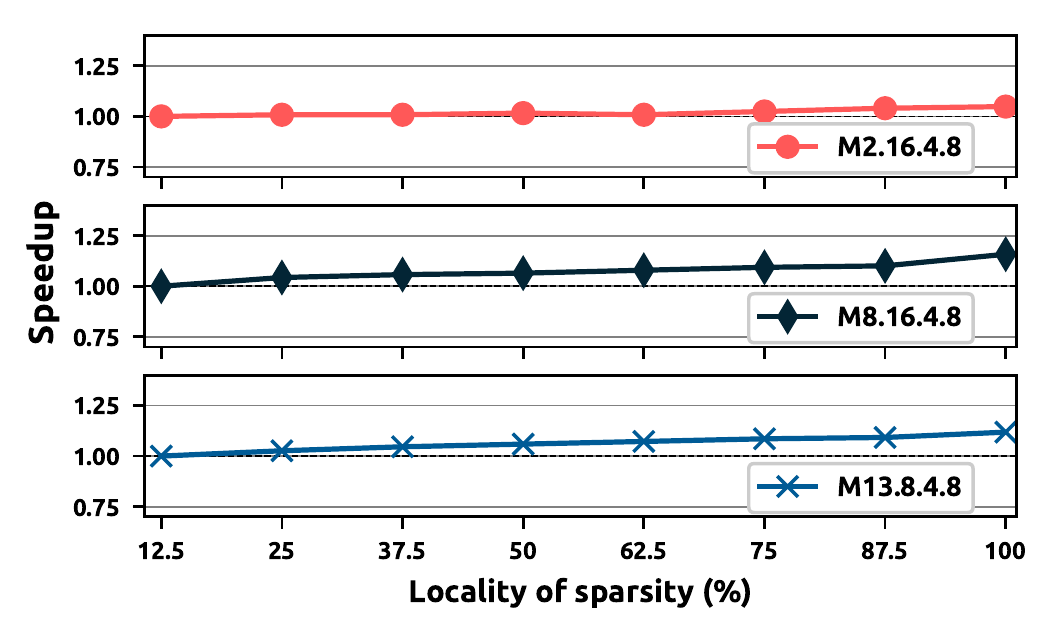}
    \vspace{-4.5mm}
   \caption{Sensitivity to locality of sparsity in SpMM.}
    \label{fig:locality-spmm}
    \vspace{-3mm}
\end{figure}

In Figures \ref{fig:locality-spmv} and \ref{fig:locality-spmm}, we illustrate the impact of the \emph{distribution} of non-zero
elements in a sparse matrix on the effectiveness of SMASH. We define a new
metric, \emph{locality of sparsity}, which is the ratio of the average number of non-zero elements
per block of the NZA to the size of each NZA block (expressed as a
percentage). A matrix with a 100\% locality of sparsity would
have no zero elements in any NZA block (i.e., all the non-zero elements are
clustered at the granularity of the NZA block size). A matrix with 12.5\% locality,
on the other hand, \kon{has exactly one non-zero element per NZA block assuming the NZA
holds 8 elements per block.}

Figures \ref{fig:locality-spmv} and \ref{fig:locality-spmm} compare the speedup of SMASH when the locality of sparsity is
varied for three different matrices ($M_{2},M_{8},M_{13}$). The results are normalized to the performance
of SMASH when the locality of sparsity is 12.5\%. The three matrices are chosen to
have widely different sparsities (0.06\%, 0.85\%, and 4.97\%). 
\vspace{-3mm}

We make two observations. First, the speedup of SMASH increases with an
increase in locality of sparsity (by up to 25\% in $M_{13}$ for SpMV when going from 12.5\% to 100\% locality of sparsity). This is
because the NZA blocks contain fewer zeros when locality is higher, and this
leads to fewer unnecessary computations on zero elements and faster scans of
bitmaps during indexing. Second, the performance impact of locality of sparsity
depends on the number of non-zero elements in the matrix, i.e., sparsity. The
benefits of locality diminish when the matrix is more sparse. This is because
\emph{indexing} of the matrices dominates the overall computation time when the sparsity
is very low and very little time is spent on computing over the
non-zero values. As a result, reducing the amount of unnecessary computation on zero elements in the NZA
blocks does not provide significant performance benefit.

\vspace{-2mm}
\subsection{Graph Applications}
\label{sec:eval_gr}
\begin{figure}[!h]
        \vspace{-2.5mm}
   \centering
    \includegraphics[scale = 0.8]{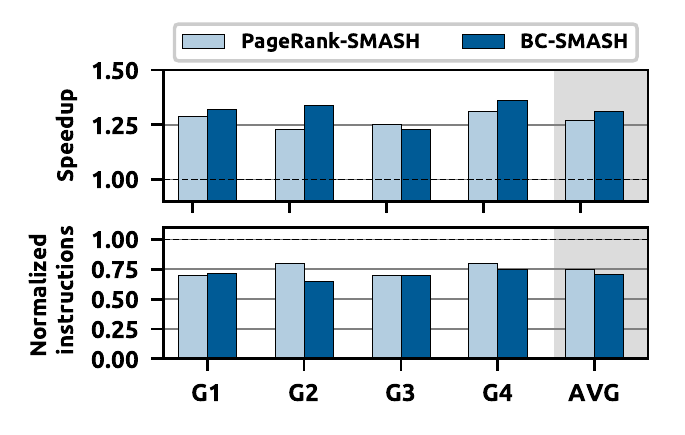}
    \vspace{-6mm}
    \caption{Speedup and normalized number of executed instructions for
	PageRank and Betweenness Centrality using SMASH (normalized to the CSR-based
	implementations).}
    \label{fig:graph}
    \vspace{-1mm}
\end{figure}

\vspace{-0.5mm}
Figure \ref{fig:graph} compares the default CSR-based and the SMASH-based implementations of
the PageRank and Betweenness Centrality applications from the Ligra benchmark
suite~\cite{ligra} in terms of performance  and the number of executed instructions.

\setlength{\textfloatsep}{2pt}

We observe that the SMASH-based implementations outperform the CSR-based
implementations by 27\% and 31\% respectively, for PageRank and Betweenness Centrality. Similar to SpMV and SpMM, SMASH's speedups in graph workloads come from executing fewer instructions to index the sparse
matrices and  avoiding the expensive pointer-chasing operations in memory. However, SMASH's
benefits are lower in graph workloads than in the SpMV and SpMM kernels since sparse matrix indexing operations in PageRank and BC 
form a smaller component of the overall execution time.  We conclude that SMASH  is effective in graph applications.

 

 \vspace{-2mm}
\subsection{Storage Efficiency}
\begin{figure}[!h]
    \centering
    \vspace{-4mm}
    \includegraphics[scale = 0.78]{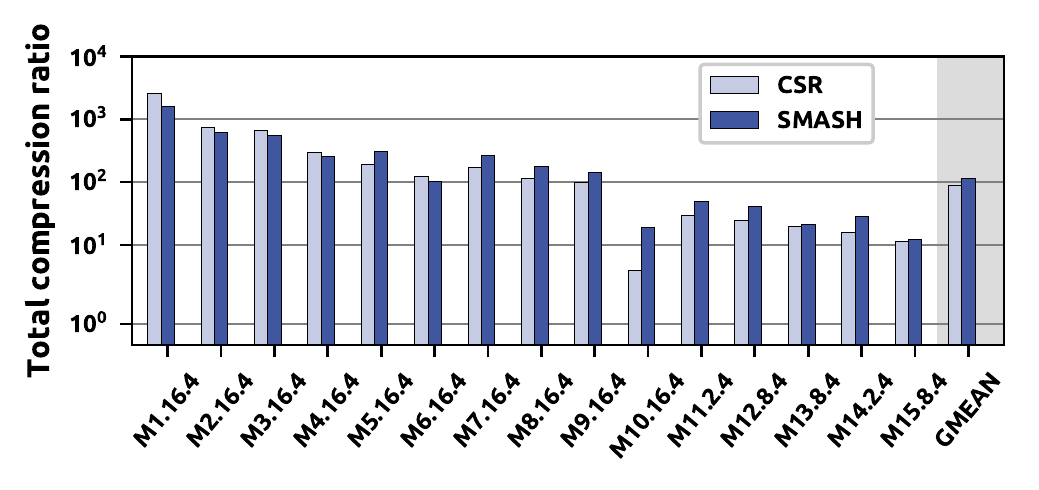}
    \vspace{-6mm}
   \caption{Total compression ratios of SMASH and CSR. Y-axis is in log scale.}
    \label{fig:storage}
    \vspace{-4mm}
\end{figure}

 \vspace{-0.5mm}
A key goal of a sparse matrix compression scheme is to efficiently store the
matrix in a manner that avoids saving zero elements and minimizes the amount of metadata required. In Figure \ref{fig:storage},
we compare the storage efficiency of SMASH and CSR by measuring the ratio
of the size of the original uncompressed matrix to the total size of all the data structures required to encode the matrix
with SMASH or the CSR format (called the \emph{total compression ratio}).

CSR stores only the non-zero elements and uses one integer per non-zero element to save the non-zero element's position in the matrix, regardless of the sparsity or the locality of sparsity of the matrix. SMASH, on the other hand, encodes each block in the NZA using a \emph{single bit} (we assume each block to hold 2 elements in Figure \ref{fig:storage}). If the non-zero elements are not contiguously located, the NZA may hold zero elements in its blocks. 


As a result, when the matrices are highly sparse, CSR only stores the few non-zero elements and their positions in the matrix.  In contrast, SMASH may store a large number of zero elements in the NZA and many zero bits in the bitmap hierarchy to encode the positions of the few non-zero elements. Hence, as we observe in Figure \ref{fig:storage}, CSR has a higher total compression ratio than SMASH for matrices that are highly sparse ($M_{1}$-$M_{4}$). 

However, as the matrices become denser, SMASH provides either a similar compression ratio to or a much higher compression ratio than CSR. The reason is twofold. First, the non-zero elements are more likely to be located close to each other at higher matrix densities (i.e., the locality of sparsity is higher). SMASH, as a result, is likely to unnecessarily store fewer zero elements in the NZA. Second, at higher matrix densities, the cost of storing one additional integer index per non-zero element in the CSR format becomes higher than encoding the zero and non-zero regions of the sparse matrix using bitmaps. Hence, in Figure \ref{fig:storage}, we observe that SMASH generally has a significantly higher (up to 2.48x) total compression ratio than CSR for matrices with higher densities. In some cases (e.g., $M13$, $M15$), even though the matrix has high density, SMASH does not provide a greatly higher compression ratio than CSR. This is because these matrices have low locality of sparsity, which causes SMASH to store more zero elements in the NZA and more zero bits in the bitmap hierarchy.

We conclude that SMASH provides high compression ratios when encoding sparse matrices, enabling highly-efficient storage of sparse matrices in memory. The hierarchical bitmap structure is highly effective in exploiting any locality of sparsity in the matrix to provide even higher compression ratios.

 \vspace{-1.7mm}

\subsection{Format Conversion Overhead}
\begin{figure}[!h]
    \centering
    \vspace{-4mm}
    \includegraphics[scale=0.71]{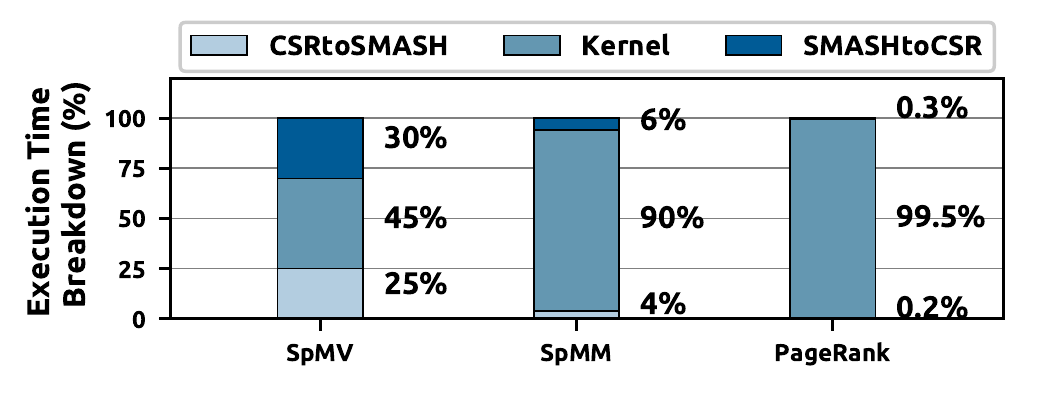}
    \vspace{-6mm}
   \caption{Breakdown of end-to-end execution time, assuming the matrix has to be stored in CSR format but processed using SMASH.}
    \label{fig:csrtosmash}
    \vspace{-3.7mm}
\end{figure}

 \vspace{-0.5mm}
In cases where the sparse matrix is already stored in another format (such as
CSR), it first needs to be converted to the hierarchical bitmap encoding in order to leverage the
indexing benefits of SMASH. Figure \ref{fig:csrtosmash} depicts the total time
spent on such conversion operations relative to the computation itself for SpMV, SpMM,
and PageRank, assuming that the sparse matrix has to be stored in the CSR format but processed using SMASH. In SpMM and PageRank, the total time spent on conversion from CSR to
SMASH and vice versa is only a small fraction of the overall computation (10\% in
SpMM and 0.5\% in PageRank), and hence imposes only a small conversion overhead.
SpMV, on the other hand, is a short-running kernel and hence, the conversion
time dominates SpMV's total execution time (55\% of the overall execution time). 

We conclude that the conversion overhead is negligible compared to the
demonstrated benefits of SMASH for long running applications and cases where
SpMV and SpMM are invoked multiple times (e.g., in neural networks and graph
applications). However, for short-running kernels that are invoked only once,
the benefits of SMASH may not justify the conversion overhead from another
format, assuming that other format is necessary for storage.
%
%
%
%
%

\vspace{-1.7mm}
\subsection{ SMASH Area Overhead}
\setstretch{0.96}
\vspace{-1mm}
The major area overhead in SMASH comes from the buffers and the registers in the BMU. Assuming
a BMU with 4 groups of 3 bitmap buffers (where the size of each buffer is 256
bytes), the total additional SRAM required for the bitmaps is 3KB and 140 bytes
for the registers. 
We evaluate the area overhead of the BMU using CACTI 6.5 \cite{CACTI}. In a
modern Intel Xeon E5-2698 CPU core, with a 32 KiB L1, 256 KiB L2 and 2.5 MiB L3 Cache,
the BMU incurs an area overhead of at most 0.076\%. 
\setstretch{0.99}

\vspace{-1mm}

\section{Related Work}
\vspace{-0.7mm}
\setstretch{0.96}

To our knowledge, this is the first work to propose a hardware-software cooperative compression scheme and indexing mechanism that 1) enables highly-compressed storage of sparse matrices, 2) provides highly-efficient hardware-supported indexing of sparse matrices to accelerate sparse matrix kernels, and 3) is general and applicable to any sparse matrix operation and matrices with any sparsity and structure. 

Sparse matrix operations have been extensively studied in the past. Prior work in this area falls into three categories: 1) sparse matrix compression formats; 2) software optimizations for sparse matrix kernels; and 3) hardware accelerators and hardware-software cooperative mechanisms for sparse matrix kernels. In this section, we briefly discuss these prior works and contrast them with SMASH.

\textbf{Sparse Matrix Compression Formats.}
Prior works propose a range of compression formats for sparse matrices~\cite{datacompression2006,im1999optimizing,formatSurvey2016,csr52015,variableblock2005,blockoptim1999,Sengupta2007scan,bccoo2014}, including the state-of-the-art formats such as CSR~\cite{csr52015}, CSR5~\cite{csr52015}, and BCSR~\cite{im1999optimizing} that are widely used today. These formats are designed to be highly-efficient in storage and can be generally applied to any sparse matrix. However, as we demonstrate in this paper, these formats incur significant performance overheads as a result of inefficient indexing. We quantitatively compare \myname{} to CSR~\cite{formatSurvey2016} and BCSR~\cite{formatSurvey2016} in Section~\ref{sec:evaluation} and demonstrate that SMASH's  hierarchical bitmap compression mechanism, along with its hardware support for indexing, significantly outperforms these formats by enabling highly-efficient storage and indexing of sparse matrices.  

Several sparse matrix compression formats~\cite{uniquekourtis2008,csxkourtis2011,smatautotune2013,csb2009,pattern2009,cocktail2012,csradaptgpus2014} leverage specific characteristics of the sparse matrix to enable more efficient storage and indexing. 
For example, CSX~\cite{csxkourtis2011} first processes each matrix to detect patterns such as dense diagonals or repeated values and then encodes them using compression formats tailored to the detected pattern. CSR-DU~\cite{uniquekourtis2008} and CSR-VI ~\cite{uniquekourtis2008} leverage repetition or similarity among non-zero values to compress the sparse matrix more efficiently. 
These works have three major shortcomings. First, they are limited in applicability to matrices that possess specific patterns in sparsity. SMASH, in contrast, can be used to compress \emph{any} sparse matrix and accelerate any operation on it. Second, these approaches require expensive preprocessing of matrices to identify patterns in sparsity. Third, similar to the general compression formats, these approaches incur significant performance overheads due to inefficient indexing. 

\textbf{Software Optimizations for Sparse Matrix Kernels.} There is a large body of prior work on software optimizations to accelerate sparse matrix computation by making sparse matrix kernels more memory or cache efficient ~\cite{caching1998,AkbudakA2017exploiting,Nishtala2007when,rbandwidth2011} and more amenable towards parallelization  ~\cite{kng2013,Buluc2008challenges,intel-mkl,Merrill2016}. A range of prior works also include general software \emph{frameworks}, such as MKL~\cite{intel-mkl} and TACO~\cite{tacolib}, that leverage these optimizations to produce more efficient code for CPUs~\cite{intel-mkl,tacolib,mergespmv2016,rbandwidth2011,henon2002pastix,bulucc2012parallel,manycoreathena2017,white1997improving,toledo1997improving,openblas1,williams2007optimization,Mellor2004optimizing,Im2004sparsity,Nishtala2007when,sparsex2018}, GPUs \cite{gpuframe2014,prspmv2010,spmmv2018,bccoo2014,gpusm2015,solversGPU,gpuopt2015}, or heterogeneous CPU-GPU systems~\cite{gpucpu2015,gpufram22012}. 
SMASH is orthogonal to these software optimizations and can be seamlessly integrated into software frameworks that employ such optimizations to further improve the performance and energy efficiency of sparse matrix computation.

\textbf{Hardware Accelerators and Hardware-Software Cooperative Mechanisms for Sparse Matrix Kernels.}
Prior works propose a range of hardware accelerators~\cite{hardanalytics2016,outerspace2018,camspm2017,Mishra2017fine,Nurvitadhi2015asparse,cambricon1,cambricon2} or FPGA designs~\cite{Yu2013design,spmvfpgacolumnmajor2014,spmvfpga2015,bwfpga2014} for sparse matrix computation. Several of these works also leverage emerging memory technologies such as memristors~\cite{memristor2016} and 3D-stacked memories~\cite{Zhu2013accelerating} to accelerate sparse matrix kernels. These prior approaches are either 1) only applicable to certain applications, such as SpMV~\cite{spmvfpga2015} or sparse neural networks~\cite{cambricon1,cambricon2}, or 2) assume hardware dedicated to a specific type of sparse matrix kernel only. SMASH, in contrast, can be generally applied across a diverse set of sparse matrix operations and does \emph{not} require fully-dedicated hardware to any particular matrix/kernel type. The hierarchy of bitmaps in SMASH can be used purely in software \emph{without} hardware support in any system, and the hardware unit proposed in SMASH can be added with low overhead to \emph{general-purpose} processors such as CPUs and GPUs.

Prior works propose a range of hardware-software cooperative mechanisms~\cite{tesseract2015,pei2015,hotpads,tsai2,hats2018,xmem,localgpu,pageover2015,tetris2017,graphP,graphR,googleworkloads2018,impica2016,graphq2019} to accelerate memory-bound operations and can be applied to accelerate sparse matrix computations. These approaches are designed to be generally applicable to a very wide range of applications. SMASH, in contrast, addresses the challenges of designing a hardware-software cooperative mechanism for sparse matrix computation. Hence, SMASH is largely orthogonal to these mechanisms and can be integrated into them to accelerate sparse matrix computation.
\setstretch{0.99}

\vspace{-2mm}
\section{Conclusion}
\vspace{-0.5mm}

\setstretch{0.95}

We introduce SMASH, a general hardware-software cooperative mechanism that accelerates sparse matrix operations and enables highly-efficient indexing and storage of sparse matrices in memory. The key idea of SMASH is to explicitly enable the hardware to recognize and exploit data sparsity. To this end, we develop a flexible and efficient sparse matrix encoding, based on a hierarchy of bitmaps, that is recognized by both hardware and software. This encoding enables efficient compression of any sparse matrix, regardless of the structure of its sparsity. We develop a hardware mechanism that can directly interpret sparse matrices encoded with hierarchical bitmaps and accelerate computation on those matrices. The bitmap representation, along with the hardware support, greatly reduces the performance overheads of the expensive indexing operations that make state-of-the-art sparse matrix formats inefficient in computation. The expressive SMASH ISA provides programmability of the hardware support and generality across a wide variety of sparse matrix kernels.  Our evaluation over a diverse set of 15 matrices and four graphs demonstrates that \myname{} significantly improves the performance of SpMV, SpMM, and two graph algorithms (PageRank and Betweenness Centrality), compared to the state-of-the-art CSR format. We believe that the new ideas introduced in SMASH are applicable beyond CPUs and can be a good fit for GPUs, hardware accelerators, and processing in/near memory engines.




\section*{Acknowledgments}

\setstretch{0.89}
We thank the anonymous reviewers for feedback. We thank the
SAFARI Research Group members for feedback and the stimulating
intellectual environment they provide. We acknowledge the generous gifts provided by our industrial partners: Alibaba, Facebook,
Google, Huawei, Intel, Microsoft, and VMware. This research was
supported in part by the Semiconductor Research Corporation. Christina Giannoula is funded for her postgraduate studies from the General Secretariat for Research and Technology (GSRT) and the Hellenic Foundation for Research and Innovation (HFRI).
\setstretch{1.0}

\newpage


\bibliographystyle{IEEEtranS.bst}
\bibliography{refs}

\end{document}